%% file: ANA-EXOT-2022-38-PAPER.tex
\pdfoutput=1
\documentclass[PAPER, atlasdraft=false, UKenglish,cernpreprint, texmf, orcidlogo]{atlasdoc}
 
\usepackage{atlaspackage}
\usepackage{bm}
\usepackage{multirow}
\usepackage{atlasbiblatex}
 
\usepackage{atlasphysics}

\graphicspath{{logos/}{figures/}}
 
\usepackage{ANA-EXOT-2022-38-PAPER-defs}

\AtlasTitle{Combination of searches for heavy spin-1 resonances using 139~fb$^{-1}$ of proton--proton collision data at $\sqrt{s} = 13$~TeV with the ATLAS detector}

\AtlasAbstract{
A combination of searches for new heavy spin-1 resonances decaying into different pairings of $W$, $Z$, or Higgs bosons, as well as directly into leptons or quarks, is presented. The data sample used corresponds to 139~\ifb\ of proton--proton collisions at $\sqrt{s}$ = 13~TeV collected during 2015--2018 with the ATLAS detector at the CERN Large Hadron Collider.
Analyses selecting quark pairs ($qq$, $bb$, $t\bar{t}$, and $tb$) or third-generation leptons ($\tau\nu$ and $\tau\tau$) are included in this kind of combination for the first time.
A simplified model predicting a spin-1 heavy vector-boson triplet is used. Cross-section limits are set at the 95\% confidence level and are compared with predictions for the benchmark model. These limits are also expressed in terms of constraints on couplings of the heavy vector-boson triplet to quarks, leptons, and the Higgs boson. The complementarity of the various analyses increases the sensitivity to new physics, and the resulting constraints are stronger than those from any individual analysis considered. The data exclude a heavy vector-boson triplet with mass below 5.8~\TeV in a weakly coupled scenario, below 4.4~\TeV in a strongly coupled scenario, and up to 1.5~\TeV in the case of production via vector-boson fusion.
}

\AtlasRefCode{EXOT-2022-38}
 
\PreprintIdNumber{CERN-EP-2024-039}

\HepDataRecord{https://www.hepdata.net/record/146753}
 
\AtlasJournalRef{\JHEP 04 (2024) 118}
\AtlasDOI{10.1007/JHEP04(2024)118}

\hypersetup{pdftitle={ATLAS document},pdfauthor={The ATLAS Collaboration}}
 
\begin{document}
 
\maketitle
 
\tableofcontents

% The next lines are included from the .//sections/1-intro.tex input file
\section{Introduction}
 
The search for heavy spin-1 resonances is an important part of the physics programme
at the Large Hadron Collider (LHC) and has been the focus of an intense effort to
uncover physics beyond the Standard Model (SM) in a broad range of final states.
In the case where new heavy gauge bosons would result from an extension
of the SM gauge group, it is possible to systematically classify them according to
their quantum numbers under the SM gauge group and to parameterize them in terms
of their masses and their couplings to SM fermions and bosons~\cite{delAguila:2010mx}.
 
One interesting possibility in this classification is the case of an iso-triplet
$\mathcal{W}$ of colourless vector bosons with zero hypercharge, which gives rise
to the Heavy Vector Triplet (HVT) framework~\cite{deBlas:2012qp,Pappadopulo:2014qza}.
This leads to a set of nearly degenerate charged, \wp$^{\pm}$, and neutral, \zp, states collectively denoted by \vp.
The charged states are denoted by \wp\ in the remainder of this article. The HVT framework allows the exploration of different coupling strengths of those
states to quarks, leptons, vector bosons, and Higgs bosons with the following interaction Lagrangian:
\begin{equation*}
\mathcal{L}_\mathcal{W}^\mathrm{int} =  - g_q \mathcal{W}^a_\mu \bar{q}_k\gamma^\mu \frac{\sigma_a}{2} q_k  - g_\ell \mathcal{W}^a_\mu \bar{\ell}_k\gamma^\mu \frac{\sigma_a}{2} \ell_k - g_H \left(\mathcal{W}^a_\mu H^\dagger \frac{\sigma_a}{2} i D^\mu H   + \mathrm{h.c.}\right),
\end{equation*}
where $q_k$ and $\ell_k$ represent the left-handed quark and lepton doublets for fermion generation $k$ ($k = 1, 2, 3$); $H$ represents the Higgs doublet; $\sigma_a$ ($a = 1, 2, 3$) are the Pauli matrices; and $g_q$, $g_\ell$, and $g_H$ correspond to the coupling strengths between the triplet field $\mathcal{W}$ and the quark, lepton, and Higgs fields, respectively.\footnote{The coupling constants $g_H$, $g_f$, $g_q$, and $g_\ell$ are used in this article. They are related to those in Ref.~\cite{Pappadopulo:2014qza} as follows: the Higgs coupling $g_H = g_V c_H$ and the universal fermion coupling $g_f = g^2 c_F / g_V$, where $g$ is the SM SU(2)$_\text{L}$ gauge coupling, while the $c$ parameters and the coupling $g_V$ are defined in Ref.~\cite{Pappadopulo:2014qza}. Couplings specific to quarks and leptons are given by $g_q = g^2 c_q / g_V$ and $g_\ell = g^2 c_\ell / g_V$.}  Right-handed fermions do not participate in these interactions. In this article, $g_q$ denotes the generation-universal quark coupling, and similarly $g_{\ell}$ denotes the generation-universal lepton coupling. The triplet field interacts with the Higgs field and thus with the longitudinally polarized $W$ and $Z$ bosons by virtue of the equivalence theorem~\cite{Cornwall:1974km,Gounaris:1986cr,Chanowitz:1985hj}. In this framework, the branching fractions for the decays $W^\prime \rightarrow WZ$, $W^\prime \rightarrow WH$, $Z^\prime \rightarrow WW$, and $Z^\prime \rightarrow ZH$, are equal for \vp\ masses above 1.5~\TeV, and other neutral diboson final states are either suppressed or forbidden. The \wp\ and \zp\ masses are taken to be degenerate, which is a very good approximation in the region of parameter space studied.
 
Although no significant excess of events has been observed to date, strong constraints have been placed
on the production of such new heavy particles. Individual analyses only constrain a subset of the coupling parameters or
have limited sensitivity to them, but a combination of searches leads to stronger simultaneous constraints, exploiting their complementarity.
Combinations of previous searches for the production of heavy resonances were performed using proton--proton (\pp) collisions at a centre-of-mass energy
$\rts = 13$~\TeV\ corresponding to an integrated luminosity of 36~\ifb, and published by
the ATLAS~\cite{EXOT-2017-31}
and CMS~\cite{CMS-B2G-18-006} Collaborations. The searches in those combinations included a heavy resonance
decaying into $VV$ or $VH$ (with $V = W$ or $Z$, and $H$ representing the SM Higgs boson) and their subsequent final states,
as well as direct decays into $\ell\ell$ and $\ell\nu$, where the charged leptons $\ell$ are either electrons or muons.
 
The ATLAS combination presented in this article
is updated to include the full
\RunTwo dataset from the LHC, constituting a sample with an integrated luminosity of 139~\ifb.
The specific searches combined in this article are those performed in
the $VV$ channels:
$WZ \to qqqq$~\cite{HDBS-2018-31e},
$qq \nu\nu$~\cite{HDBS-2018-10},
$\ell\nu qq$~\cite{HDBS-2018-10},
$qq \ell\ell$~\cite{HDBS-2018-10},
$\ell\nu\ell\ell$~\cite{ATLAS:2022zuc},
$WW \to qqqq$~\cite{HDBS-2018-31e},
$\ell\nu qq$~\cite{HDBS-2018-10};
the $VH$ channels:
$WH \to qq bb$~\cite{HDBS-2018-11}, $\ell\nu bb$~\cite{ATLAS:2022enb}, and
$ZH \to qq bb$~\cite{HDBS-2018-11}, $\nu\nu bb$~\cite{ATLAS:2022enb},
$\ell\ell bb$~\cite{ATLAS:2022enb};
and the lepton--antilepton channels:
$\ell\ell$~\cite{EXOT-2018-08}, $\ell\nu$~\cite{EXOT-2018-30}.
New channels added to this combination are $\tau\nu$~\cite{ATLAS:taunu}, $\tau\tau$~\cite{HDBS-2018-46},
and the diquark channels: $q\bar{q}$~\cite{EXOT-2019-03} (written as \enquote{$qq$}), $b\bar{b}$~\cite{EXOT-2019-03} (written as \enquote{$bb$}),
$t\bar{t}$ $\rightarrow$ $qqqqbb$ (fully hadronic), named \enquote{tt0L}~\cite{EXOT-2018-48}, $tb$ $\rightarrow$ $qqbb$ (fully
hadronic), named \enquote{tb0L}~\cite{ATLAS:2023ibb}, and $tb$ $\rightarrow$ $\ell\nu bb$ (semi-leptonic), named \enquote{tb1L}~\cite{ATLAS:2023ibb}.
The CMS Collaboration has also performed dedicated searches in many of these final states with the full \RunTwo dataset~\cite{CMS:2022pjv,CMS:2021xor,CMS:2021klu,CMS:2021fyk,CMS:2021itu,CMS:2017dtg,CMS:2021ctt,CMS:2022krd,CMS:2022ncp,CMS:2016xbv,CMS:2019gwf,CMS:2022eud,CMS:2018rkg,CMS:2021mux,CMS:2023gte}.
In this article, the $VV$ and $VH$ decay channels are collectively called \enquote{bosonic}, the lepton--antilepton decay channels are collectively called \enquote{leptonic}, and the diquark channels are collectively called \enquote{quarkonic}. The analyses generally search for narrow resonances in the final-state mass distribution, with the signal's shape extracted from Monte Carlo (MC) simulation of specific models. The background's shape and normalization are extracted from a combination of MC simulation and functional fits to the data, often relying on dedicated control regions to extract the various background contributions. The mass distributions and associated systematic uncertainties from the various channels are combined using a likelihood method which takes correlations into account.

% End of text imported from the .//sections/1-intro.tex input file

% The next lines are included from the .//sections/2-models.tex input file
\section{Signal benchmark scenarios}
\label{sec:models}
 
\subsection{Benchmark points: HVT model A, B and C}
 
Among the variety of generic models that can be constrained in the HVT framework, three explicit scenarios are used as benchmarks for interpretation of the results.
The first two benchmarks correspond to quark--antiquark annihilation ($q\bar{q}$) production processes, while the third benchmark
is designed to focus solely on the rare vector-boson fusion (VBF) process.
 
For the $q\bar{q}$ processes, the two scenarios emphasize the relative strengths of the $g_H$ and $g_f$ couplings. The first scenario, referred to as model A, reproduces the phenomenology of weakly coupled models based on an extended gauge symmetry~\cite{Barger:1980ix}. In this case, the couplings are $g_H = -0.56$ and $g_f = -0.55$, with the universal fermion coupling $g_f = g_q = g_{\ell}$. The second $q\bar{q}$ scenario, referred to as model B, implements a strongly coupled scenario as in composite Higgs models~\cite{Contino:2011np} with $g_H = -2.9$ and $g_f = 0.14$.\footnote{In terms of the coupling constants in the notation of Ref.~\cite{Pappadopulo:2014qza}, the choices for models A and B correspond to $g_V = 1$ and $g_V = 3$, respectively.} In model B, the $V^\prime$ resonances are broader than in the weakly coupled scenario, model A, but remain narrow relative to the experimental resolution. The relative width of the  $V^\prime$ resonances, $\Gamma / m$, is below 5\% over much of the parameter space explored.
 
The $VV$ and $VH$ analyses are designed to optimize the signal selection efficiency in their respective final states. However, signal decays into both $VV$ and $VH$ are included in each diboson analysis to capture the full acceptance of the signals. The acceptance in individual channels is the same for models A and B. The relative branching fractions are set to the appropriate contributions for each. For masses below 800~\GeV, model B is not compatible with SM precision measurements, due to larger mixing between the SM gauge bosons and the heavy vector resonance. Therefore, this region is not considered in the combination.
 
For the $q\bar{q}$ process with decay of the $V^\prime$ into fermion--antifermion final states, branching fractions are larger in model A: the branching fraction for decay into the lepton--antilepton final state is approximately 4\% for each generation, whereas it is only about 0.2\% in model B. In contrast, the branching fractions for decays into $VV$ or $VH$ are about 2\% each in model A, whereas they are close to 50\% each in model B.
 
The VBF production scenario is referred to as model C. In this case, the $V^\prime$ resonance's
couplings are set to $g_H = 1$ and $g_f = 0$. Therefore, Model C is in a phase-space
domain separate from that of model A or B, and assumes no $q\bar{q}$ production.
 
\subsection{Benchmark HVT coupling planes}
\label{subsec:hvtcoupling}
 
The interpretations can be extended beyond the benchmark points described above by considering the
two-dimensional coupling-parameter space. The different production mechanisms and
decay modes provide sensitivity to different regions of the parameter space,
with production via $q\bar{q}$ and quarkonic final states providing sensitivity to $g_q$, while
VBF production provides sensitivity to $g_q=0$.
Likewise, decays into leptonic final states provide sensitivity to $g_\ell$, and
decays into bosonic final states provide sensitivity to $g_H$.
Five coupling planes are used, and are described below.
 
Previous combinations~\cite{EXOT-2017-31, CMS-B2G-18-006} focused on coupling planes such as
$\{g_H, g_f\}$ (assuming fermion universality)
and  $\{g_q, g_{\ell}\}$ with $g_H=-0.56$.
The former coupling plane includes
HVT model A and B benchmark points, while the latter is based on the HVT model A benchmark point,
due to the $g_H$ coupling requirement.
 
The inclusion of channels with third-generation fermions allows constraints to be placed on previously unexplored parameters
such as a specific third-generation quark coupling ($g_{q3}$) or third-generation lepton
coupling ($g_{\ell3}$). 
Such results are particularly interesting because of recent measurements that hint at a possible
violation of lepton-flavour universality in rare $B$-meson semi-tauonic
decays~\cite{BaBar:2013mob,LHCb:2015gmp,Belle:2015qfa,Belle:2017ilt,LHCb:2017smo,LHCb:2017rln,Belle:2019rba,LHCb:2017vlu,LHCb:2022piu,LHCb:2023zxo}.
 
Two new coupling planes are constructed to test the HVT model A benchmark point:
\begin{itemize}
\item the $\{g_H, g_{q3}\}$ coupling plane has the additional requirement $g_{\ell}=g_{q12}=-0.55$, where $g_{q12}$ is the coupling to first- and second-generation quarks;
\item the $\{g_{q3}, g_{\ell 3}\}$ coupling plane has the additional requirements $g_{\ell 12}=g_{q12}=-0.55$
and $g_H=-0.56$, where $g_{\ell 12}$ is the coupling to first- and second-generation leptons.
\end{itemize}
The $\{g_H, g_{q3}\}$ coupling plane probes a specific region of the model's phase space where the bosonic coupling is strong, as well as the third-generation quark coupling. The $\{g_{q3}, g_{\ell 3}\}$ coupling plane explores the scenario where the third-generation quark and lepton couplings vary relative to each other and can be dominant or suppressed in comparison with the other couplings.
 
A third new coupling plane, $\{g_{q12}, g_{q3}\}$, focuses on the couplings to quarks only
and sets all other couplings to zero, i.e.\ $g_{\ell}=0$ and $g_H=0$. This plane probes boson-phobic and lepto-phobic new phenomena, but with non-universal couplings between the first/second-generation quarks and third-generation quarks.

% End of text imported from the .//sections/2-models.tex input file

% The next lines are included from the .//sections/3-strategy.tex input file
\section{Combination strategy}
\label{sec:strategy}
 
The strategy is to consider orthogonal ATLAS analyses which were used to search independently for new phenomena in specific final states, and combine them to obtain more stringent constraints on the relevant parameters of the models described above. After a search phase in which local $p$-values are extracted, the results are presented in two formats: one-dimensional (1D) and two-dimensional (2D) exclusion limits. The 1D format provides limits on the cross-section times branching fraction (\sigbr) of the signal process versus the resonance pole mass. The 2D limits display constraints on various pairs of coupling parameters, expressed as exclusion contours for a subset of resonance masses.  These exclusion contours provide a more model-independent interpretation.
 
To avoid using the same data for different final states, the analyses optimized for the best sensitivity as stand-alone searches are each evaluated for orthogonality.
If overlap between analyses is found, minimal additional requirements are applied to achieve orthogonality.
In general, this involves requiring the exclusivity of final states as described in Table~\ref{tab:selection} of Section~\ref{sec:selection}.
If overlaps remain, complementary kinematic cuts are applied to variables used by the overlapping analyses (such as the jet mass) to orthogonalize them.
These requirements are chosen to maximize the sensitivity of the combined result.
In cases where achieving complete orthogonality is not feasible, the level of overlap is assessed to determine whether it has a negligible impact (below 5\%) on the cross-section limits.
If this is the case, then the overlap is ignored; however, if the overlap has a non-negligible impact on the result, then it is accounted for.
In this scenario, an analysis priority order is established to preferentially remove, on an event-by-event basis, the overlapping data events
from the discriminating variable distribution of the lower-priority analysis while keeping them in the higher-priority analysis.
Which analysis has higher or lower priority depends on the region of phase space being explored and whether the removal of the events would invalidate the signal or background estimation in either analysis. For example, it is checked that the removal of the overlapping events does not modify the validity of the fit to data. This is achieved by performing the same fit before and after overlap removal and checking that the behaviour of the fit result is the same. In all cases the change to the fit was negligible as a function of the discriminating variable (within 5\%). The new fits are then used for completeness and no further systematic uncertainty is applied. Furthermore, overlap removal is only applied to the signal regions such that each individual analysis considers its own control and validation regions, and their normalization is not correlated across analyses involved in the combination.\footnote{No significant pulls in normalizations appeared when performing the full combination.}

% End of text imported from the .//sections/3-strategy.tex input file

% The next lines are included from the .//sections/4-detector.tex input file
\section{ATLAS detector}
 
\newcommand{\AtlasCoordFootnote}{
ATLAS uses a right-handed coordinate system with its origin at the nominal interaction point (IP)
in the centre of the detector and the \(z\)-axis along the beam pipe.
The \(x\)-axis points from the IP to the centre of the LHC ring,
and the \(y\)-axis points upwards.
Polar coordinates \((r,\phi)\) are used in the transverse plane,
\(\phi\) being the azimuthal angle around the \(z\)-axis.
The pseudorapidity is defined in terms of the polar angle \(\theta\) as \(\eta = -\ln \tan(\theta/2)\).
Angular distance is measured in units of \(\Delta R \equiv \sqrt{(\Delta\eta)^{2} + (\Delta\phi)^{2}}\).}
 
The ATLAS experiment~\cite{PERF-2007-01} at the LHC is a multipurpose particle detector
with a forward--backward symmetric cylindrical geometry and a near \(4\pi\) coverage in
solid angle.\footnote{\AtlasCoordFootnote}
It consists of an inner tracking detector surrounded by a thin superconducting solenoid
providing a \qty{2}{\tesla} axial magnetic field, electromagnetic and hadronic calorimeters, and a muon spectrometer.
The inner tracking detector covers the pseudorapidity range \(|\eta| < 2.5\).
It consists of silicon pixel, silicon microstrip, and transition radiation tracking detectors.
Lead/liquid-argon (LAr) sampling calorimeters provide electromagnetic (EM) energy measurements
with high granularity within the region \(|\eta|< 3.2\).
A steel/scintillator-tile hadronic calorimeter covers the central pseudorapidity range (\(|\eta| < 1.7\)).
The endcap and forward regions are instrumented with LAr calorimeters
for EM and hadronic energy measurements up to \(|\eta| = 4.9\).
The muon spectrometer surrounds the calorimeters and is based on
three large superconducting air-core toroidal magnets with eight coils each.
The field integral of the toroids ranges between \num{2.0} and \qty{6.0}{\tesla\metre}
across most of the detector.
The muon spectrometer includes a system of precision tracking chambers up to \(|\eta| = 2.7\) and fast detectors for triggering up to \(|\eta| = 2.4\).
The luminosity is measured mainly by the LUCID--2~\cite{LUCID-2} detector, which is located close to the beampipe.
A two-level trigger system is used to select events~\cite{TRIG-2016-01}.
The first-level trigger is implemented in hardware and uses a subset of the detector information
to accept events at a rate below \qty{100}{\kHz}.
This is followed by a software-based trigger that
reduces the accepted event rate to \qty{1}{\kHz} on average
depending on the data-taking conditions.
A software suite~\cite{ATL-SOFT-PUB-2021-001} is used in data simulation, in the reconstruction
and analysis of real and simulated data, in detector operations, and in the trigger and data acquisition
systems of the experiment.

% End of text imported from the .//sections/4-detector.tex input file

% The next lines are included from the .//sections/5-datamc.tex input file
\section{Data and Monte Carlo simulation}
 
The data sample was collected by the ATLAS detector during the $pp$ collision running of the LHC at $\sqrt{s} = 13$~\TeV from 2015 to 2018. Events were selected for the different channels with various triggers, as described in their respective publications. Channels featuring charged or neutral leptons were selected with single or multiple electron and muon triggers with various \pT\ thresholds and isolation requirements, or with missing transverse momentum (\met) triggers with thresholds that were raised once the LHC's peak luminosity had increased significantly. A high-\pt\ jet trigger was used in the fully hadronic channels. More details of the event selection for each analysis are provided in Section~\ref{sec:selection}.
Only events taken during stable beam conditions and satisfying detector- and data-quality requirements~\cite{DAPR-2018-01}, which include the calorimeters, muon system and inner tracking detectors being in normal operation, are considered. After all of these quality criteria, the total integrated luminosity amounts to 139~\ifb.
 
The interpretation in this combination relies on MC simulation to model the shape and normalization of the signals described in Section~\ref{sec:models}. Signal events for the HVT (\zp, \wp) for all three models were generated with \MGNLO[2.6.5]~\cite{madgraph} at leading order (LO) using the \NNPDF[2.3lo] parton distribution function (PDF) set~\cite{Ball:2012cx}. The generated events were interfaced to \PYTHIA[8.240]~\cite{pythia8} for parton showering, hadronization, and the underlying event.  \PYTHIA used the A14 set of tuned parameters~\cite{ATL-PHYS-PUB-2014-021} for events generated at LO with \MGNLO. Interference between the signal events and SM processes was not taken into account because this combination assumes that the narrow-width approximation is valid. The range of cross-sections and branching ratios used for the various channels in the combination at different points in the HVT parameter space are calculated at LO using the HVT calculator tool~\cite{Pappadopulo:2014qza,Baker:2022zxv}.
 
Simulated background event samples are used to estimate the main backgrounds, or to extrapolate backgrounds from control regions in the case of analyses in the $\ell\nu$, $\ell\nu\ell\ell$, tb0L, tb1L, $\tau\nu$, and $\tau\tau$ channels. In other channels, namely $\ell\ell$, $qqqq$, $qqbb$, $q\bar{q}$, $b\bar{b}$, and tt0L, the data are used to determine the normalization and/or shape of the background distributions. 
Although the production of background MC samples varies somewhat depending on the specific analysis, most MC samples were produced as follows. Diboson ($WW$, $WZ$, $ZZ$) events were generated with \SHERPA~\cite{Gleisberg:2008ta} or \POWHEGBOX~\cite{Nason:2004rx, Frixione:2007vw, Alioli:2010xd}; $W$+jets and $Z$+jets events were generated with \SHERPA for up to two partons at NLO, and up to four partons at LO, using the \OPENLOOPS~\cite{Cascioli:2011va} and Comix~\cite{Gleisberg:2008fv} programs, respectively. The production of top-quark pairs and single top quarks was performed at NLO with \POWHEGBOX. For the $\ell\nu$ channel, the dominant Drell--Yan background was modelled using \POWHEGBOX with mass-dependent corrections up to next-to-next-to-leading order in QCD and NLO in electroweak interactions. More specific details can be found in the publication for each analysis (see Table~\ref{tab:selection} for references).
 
For the $VV$ and $VH$ decay channels involving leptonic decays of vector bosons, $\tau$-leptons are included in the simulation of the signals, and the contamination from reconstructed $\tau$-lepton final states involving only electrons and muons is negligible. The impact of $\tau$-leptons contaminating other channels is very small and thus neglected.
 
For all MC samples, except those produced with \SHERPA, $b$-hadron and $c$-hadron decays were performed with \EVTGEN[1.6.0]~\cite{Lange:2001uf}. The simulated event samples include the effect of multiple $pp$ interactions per bunch crossing, as well as the effect on the detector response due to interactions from bunch crossings before or after the one containing the hard interaction. These effects are collectively referred to as \enquote{\pileup{}}. The simulation of \pileup collisions was performed with \PYTHIA[8] and tuned to reproduce the average number of \pileup interactions per bunch crossing observed in a given data-taking period. Most of the MC samples were processed through a detailed simulation of the detector response with \GEANT~\cite{geant, SOFT-2010-01}. A small subset of the MC samples were processed with a fast parameterization of the calorimeter response~\cite{ATL-PHYS-PUB-2010-013}, while \GEANT was used for the response of the other detector components. In all cases, events were reconstructed with the same software~\cite{ATL-SOFT-PUB-2021-001} as was used for the data.

% End of text imported from the .//sections/5-datamc.tex input file

% The next lines are included from the .//sections/6-selection.tex input file
\section{Event selection}
\label{sec:selection}
 
The event selection for the different analyses is summarized in Table~\ref{tab:selection} along with their experimental signatures. A full description is available in Refs.~\cite{HDBS-2018-31e,HDBS-2018-10,HDBS-2018-11,ATLAS:2022zuc,ATLAS:2022enb,EXOT-2018-30,ATLAS-CONF-2021-025,EXOT-2018-08,HDBS-2018-46,EXOT-2018-48,ATLAS:2023ibb,EXOT-2019-03}.
Special attention was given to the event selection to achieve orthogonality between the different channels, as described in Section~\ref{sec:strategy}.
The channels are broadly separated into four categories, depending on the targeted decay state of the intermediate resonance: a vector-boson pair ($VV$), a $W$ or $Z$ boson with an associated Higgs boson ($VH$), a pair of leptons, and a pair of quarks. For the $VV$ category, in the semi-leptonic and fully leptonic channels, the searches are further split into optimized selections for VBF production.
 
\begin{table}
\begin{center}
\caption{Summary of event selection for the signal regions of the analyses used in the combination. The entries which are separated by commas denote an \enquote{OR} of different signal regions. For jets, \enquote{j} indicates a small-$R$ jet, while \enquote{J} denotes a large-$R$ jet. The symbol $\subset$ denotes any combination of at least one of the objects inside the bracket up to the number shown. Entries denoted by a \enquote{-} indicate no requirement for that object. The VBF column indicates whether the analysis has an additional category for VBF events. The abbreviation \enquote{Discr.} stands for the discriminating variable used in a given search.}
\label{tab:selection}
\begin{tabular}{ccccccccc}
\toprule
Analysis & Leptons & \met & Jets & $b$-tags & Top-tags & VBF & Discr. & Reference \\
\midrule
$WW$/$WZ$ $\rightarrow$ $qqqq$ & 0 & Veto & $\ge$2J & - & - & - & $m_{VV}$ & \cite{HDBS-2018-31e} \\%\hline
$WW$/$WZ$ $\rightarrow$ $\ell\nu qq$ & 1$e$, 1$\mu$ & Yes & $\ge$2j, $\ge$1J & 0, 1, 2 & - & Yes & $m_{VV}$ & \cite{HDBS-2018-10} \\%\hline
$WZ$ $\rightarrow$ $qq \nu\nu$ & 0 & Yes & $\ge$1J & 0 & - & Yes & $m_{VV}$ & \cite{HDBS-2018-10} \\%\hline
$WZ$ $\rightarrow$ $qq \ell\ell$ & 2$e$, 2$\mu$ & - & $\ge$2j, $\ge$1J & 0 & - & Yes & $m_{VV}$ & \cite{HDBS-2018-10} \\%\hline
$WZ$ $\rightarrow$ $\ell\nu\ell\ell$ & 3 $\subset$ ($e$, $\mu$) & Yes & - & 0 & - & Yes & $m_{VV}$ & \cite{ATLAS:2022zuc} \\
\hline
$WH/ZH$ $\rightarrow$ $qqbb$ & 0 & Veto & $\ge$2J & 1, 2 & - & - & $m_{VH}$ & \cite{HDBS-2018-11} \\%\hline
$ZH$ $\rightarrow$ $\nu\nu bb$ & 0 & Yes & $\ge$2j, $\ge$1J & 1, 2 & - & - & $m_{VH}$ & \cite{ATLAS:2022enb} \\%\hline
$WH$ $\rightarrow$ $\ell\nu bb$ & 1$e$, 1$\mu$ & Yes & $\ge$2j, $\ge$1J & 1, 2 & - & - & $m_{VH}$ & \cite{ATLAS:2022enb} \\%\hline
$ZH$ $\rightarrow$ $\ell\ell bb$ & 2$e$, 2$\mu$ & Veto & $\ge$2j, $\ge$1J & 1, 2 & - & - & $m_{VH}$ & \cite{ATLAS:2022enb} \\
\hline
$\ell\nu$ & 1$e$, 1$\mu$ & Yes & - & - & - & - & $m_\mathrm{T}$ & \cite{EXOT-2018-30} \\
$\tau\nu$ & 1$\tau$ & Yes & - & - & - & - & $m_\mathrm{T}$ & \cite{ATLAS:taunu} \\
$\ell\ell$ & $\ge$2$e$, $\ge$2$\mu$ & - & - & - & - & - & $m_{\ell\ell}$ & \cite{EXOT-2018-08} \\
$\tau\tau$ & 0, 1$e$, 1$\mu$ & Yes & - & 0, $\ge$1 & - & - & $m_{\tau\tau}$ & \cite{HDBS-2018-46} \\
\hline
tt0L & 0 & - & 2J & 1, 2 & 2 & - & $m_{tt} $  & \cite{EXOT-2018-48} \\
tb0L & 0 & - & $\ge$(1j+1J) & $\ge$1 & 1 & - & $m_{tb}$ & \cite{ATLAS:2023ibb} \\
tb1L & 1$e$, 1$\mu$ & Yes & 2j, 3j & 1, 2 & - & - & $m_{tb}$ & \cite{ATLAS:2023ibb} \\
\hline
$qq$ & 0 & - & 2j & 0 & - & - & $m_{jj}$ & \cite{EXOT-2019-03} \\
$bb$ & 0 & - & 2j & 1, 2 & - & - & $m_{bb}$ & \cite{EXOT-2019-03} \\
\bottomrule
\end{tabular}
\end{center}
\end{table}
 
\clearpage

% End of text imported from the .//sections/6-selection.tex input file

% The next lines are included from the .//sections/7-systematics.tex input file
\section{Systematic uncertainties}
 
The experimental and theoretical systematic uncertainties considered across the range of analyses included in the combination are discussed qualitatively in this section, leaving the details to the original publication for each analysis. The 
uncertainties are assessed as a function of the discriminating variable in each channel in the combination. These uncertainties apply to both the signal and (simulation-based) background distributions, and are treated as either correlated 
or uncorrelated between the signal and background in each channel, as appropriate.
These systematic uncertainties weaken the upper limit on the cross-section by as much as 20\%, compared to applying only the statistical uncertainties.
 
Lepton systematic uncertainties refer specifically to electrons and muons, and comprise energy/momentum scale and resolution uncertainties, as well as those arising from lepton identification, reconstruction, isolation, and trigger efficiencies.
These uncertainties are treated as being correlated between the signal and the background estimates, and across the analysis channels where the same lepton selection criteria were used.
 
The $\tau$-lepton-related uncertainties, which are considered in the $\tau\nu$ and $\tau\tau$ analyses, are related to the determination of the $\tau$-lepton identification efficiency, reconstruction efficiency, and energy scale~\cite{ATLAS-CONF-2017-029}. The overlap removal procedure, particularly when using the number of charged particles in the final state to distinguish between $\tau$-lepton decay modes, can also lead to non-negligible uncertainties. Inaccuracies in simulating particle transport through the detector material can lead to further reconstruction efficiency uncertainties, which are also considered in the $\tau\nu$ and $\tau\tau$ analyses.
 
Systematic uncertainties in the missing transverse momentum arise mainly from the \enquote{soft term} resolution and scale, as well as the trigger used by some analyses. In all cases, these uncertainties are applied to both the signal and background distributions, and are treated as being correlated across the analysis channels.
 
For the analyses that select small-$R$ jets, which are reconstructed using the \antikt algorithm~\cite{Cacciari:2008gp} with a radius parameter of \(R=0.4\), the largest uncertainties are often from the jet energy scale and resolution, followed by uncertainties related to jet flavour, \pileup effects, punch-through, and the jet-vertex tagger~\cite{JETM-2018-05}.  As with the other experimental uncertainties discussed so far, the small-$R$ jet uncertainties are considered for both the signal and background distributions, and treated as being correlated across the analysis channels. 
For large-$R$ jets, reconstructed using the same \antikt algorithm but with a radius of \(R=1.0\), the jet energy/mass scale and resolution uncertainties are important, and so are the experimental uncertainties related to jet substructure variables, such as uncertainties in the $D_2$ scale~\cite{Larkoski:2016} and resolution. These systematic uncertainty sources are broken up into a number of orthogonal subcomponents and treated as separate nuisance parameters, so the overall uncertainty category is summarized here. This class of uncertainties is considered for both the signal and background distributions, and is correlated across the analysis channels where the same jet collections were used.
 
The flavour-tagging uncertainties are evaluated from the measured tagging efficiencies and mistag rates in various kinematic regions. The broad categories of flavour-tagging uncertainties~\cite{FTAG-2018-01,ATL-COM-PHYS-2022-246} considered are $b$-tagging, $c$-tagging, light-jet tagging, and tagging extrapolation for high-\pT regions.
 
For small-$R$-jet, large-$R$-jet, and $b$-tagging systematic uncertainties, the respective sources are treated as correlated only between the analyses that use the same collections and definitions. The uncertainty in the integrated luminosity is determined to be 1.7\%, using the LUCID-2 detector for the primary luminosity measurements~\cite{LUCID-2,ATLAS-CONF-2019-021}.
 
Dominant backgrounds differ between the various analyses, and therefore different theoretical uncertainties are often prioritized or studied in more detail for each search. For the top-quark and diboson backgrounds, which are relatively small in leptonic analysis channels, only the theory cross-section uncertainties are considered, whereas the $VV$ and $VH$ analyses also take into account the modelling uncertainties for these backgrounds. In some cases, analyses also add an uncertainty to cover extrapolations from low to high masses for certain background processes with large statistical uncertainties in the simulated samples. Because the probed phase space differs between the many analyses, and some normalizations are left to float freely in the fit performed in their respective analysis, these uncertainties are treated as uncorrelated across the channels to allow better background modelling.
 
For the multi-jet background, most analyses use a data-driven approach and consider uncertainties in both the method and modelling of any extrapolation. For the leptonic channels, the $W$+jets background uncertainty is also included in this category. These uncertainties are treated as uncorrelated across the analysis channels.
 
Most of the $VV$ and $VH$ analyses, as well as the $\tau\nu$ analysis, consider cross-section and modelling uncertainties affecting the $Z$+jets and $W$+jets background processes. For the $WW$ $\rightarrow$ $\ell\nu qq$ analysis, in which $W$+jets is a particularly important background, the PDF eigenvector uncertainty and the renormalization/factorization scale and strong interaction coupling strength (\alphas) uncertainties are also included. For the leptonic analysis channels, the Drell--Yan background is by far the largest, and thus its theoretical uncertainties are studied in great detail. As such, these analyses consider the uncertainties related to the choice of \alphas value and PDF, as well as those related to electroweak corrections and photon-induced background contributions. These Drell--Yan uncertainties were specifically studied and arranged such that they are consistent and can be correlated across the different leptonic analysis channels. The systematic uncertainties in the background modelling for the fully hadronic analysis, $qqqq$, are embedded in the fit function used to model the background.

% End of text imported from the .//sections/7-systematics.tex input file

% The next lines are included from the .//sections/8-statistics.tex input file
\section{Statistical treatment}
 
The combination of the individual channels proceeds with a simultaneous analysis of the signal discriminants across all of the channels. For each signal model being tested, only the channels sensitive to the signal hypothesis are included in the combination. The statistical treatment of the data is based on the RooFit~\cite{Verkerke:2003ir}, RooStats~\cite{Moneta:2011bQ}, and HistFactory~\cite{Cranmer:2012sba} data modelling and handling toolkits. Results are calculated in two different signal parameterization paradigms, corresponding to the 1D upper limits on the cross-section times branching fraction (\sigbr) and 2D limits on coupling strengths. The statistical treatment of each case is described below.
 
\subsection{1D upper limits}
 
In the case of 1D upper limits on \sigbr, the overall signal strength, $\mu$,
is the parameter of interest and is defined as a scale factor multiplying the cross-section times branching fraction predicted by the signal hypothesis. The combined analysis follows the frequentist approach with a test statistic based on the profile-likelihood ratio~\cite{Cowan:2010js}. This test statistic ($\mathcal{T}$) is defined as twice the negative logarithm of the ratio of the conditional (fixed-$\mu$) maximum likelihood to the unconditional maximum likelihood, each obtained from a fit to the data:
 
\begin{equation}
\label{eqn:LLRnom}
\mathcal{T}(\mu) = -2 \ln \left\{ \frac{L(\mu,\hat{\hat{\bm{\theta}}}(\mu))}{L(\hat{\mu},\hat{\bm{\theta}}(\hat{\mu}))} \right\},
\end{equation}
 
where $\hat{\mu}$ and $\hat{\bm{\theta}}$ are the maximum-likelihood estimators of the signal strength and the nuisance parameters, respectively, and
$\hat{\hat{\bm{\theta}}}$ maximizes the likelihood for the given value of $\mu$.
The best fitted signal strength, $\hat{\mu}$, is bounded from below at zero.
 
The joint likelihood, ${L}$, over all analysis channels and histogram bins is given by:
 
\begin{equation*}{L}(\mu, \theta) = \prod_{c} \prod_{i} \mathrm{Pois}\left(n^{\mathrm{obs}}_{ci} \left| \, n^\mathrm{sig}_{ci}(\mu, \vec{\theta}) + n^{\mathrm{bkg}}_{ci}( \vec{\theta}) \right. \right) \prod_{k} f_{k}(\theta_k)~,
\end{equation*}
 
where the index $c$ represents the analysis channel, $i$ represents the bin in the signal discriminant distribution, $n^{\mathrm{obs}}$ is the observed number of events, $n^\mathrm{sig}$ is the expected number of signal events, $n^\mathrm{bkg}$ is the expected number of background events, $\vec{\theta}$ denotes the vector of nuisance parameters, and $\mathrm{Pois}(x|y)$ is the Poisson probability to observe $x$ events when $y$ events are predicted.
 
The effect of a systematic uncertainty with index $k$ on the binned likelihood is modelled with an associated nuisance parameter, $\theta_{k}$, constrained with a corresponding probability density function $f_k(\theta_k)$. In this manner, correlated effects across the different channels are modelled by the use of a common nuisance parameter and its corresponding probability density function. The $f_k(\theta_k)$ terms are Poisson-distributed for bin-by-bin MC statistical uncertainties, and Gaussian-distributed for all other terms.
 
Given the large number of search channels included in the likelihood, the sampling distribution of the profile-likelihood test statistic is assumed to follow the chi-squared ($\chi^2$) distribution, and thus asymptotic formulae are used for the evolution of the likelihood as a function of $\mu$~\cite{Cowan:2010js}.  In certain instances, such as low-yield, high-mass tails of resonant mass distributions, the asymptotic approximation is expected to be less reliable.  In these cases, MC pseudo-experiments are used to assess its accuracy. The asymptotic approximation is found to lead to \sigbr\ limits that are up to 20\% stronger than those obtained with MC pseudo-experiments. Specifically, this discrepancy was prominent for the leptonic subcombination at the highest signal pole mass. However, the limits for the rest of the leptonic subcombination range, as well as for the bosonic, quarkonic, and full combination, were found to be in very good agreement (less than 5\% difference). The corresponding change in the mass limits is negligible (within rounding).
 
When evaluating limits in the HVT model with mass-degenerate \wp\ and \zp\ bosons, each of the contributing signal processes is normalized to the \sigbr\ value predicted by the HVT model couplings at the point of interest, thereby fixing the cross-section ratios $\sigma(pp \rightarrow W^\prime)/\sigma(pp \rightarrow Z^\prime)$ and the relative branching fractions of all decays. The HVT model benchmarks make a model-dependent assumption about these cross-section ratios, so the resulting upper limits cannot be directly interpreted as general limits on \sigbr.
 
Upper limits on $\mu$ for the signal models being tested at the simulated resonance masses are evaluated at the 95\% confidence level (CL) following the CL$_\text{s}$ prescription~\cite{Read:2002hq}. Lower limits on the mass of new resonances in these models are obtained by finding the maximum resonance mass where the 95\% CL upper limit on $\mu$ is less than or equal to unity. This mass is found by interpolating  between the limits on $\mu$ at the simulated signal masses. The interpolation assumes monotonic and smooth behaviour of the efficiencies for the signal and background processes, and that the impact of the variation of the shape of signal mass distributions between adjacent test masses is negligible.
 
\subsection{2D limits}
 
When calculating 1D upper limits on \sigbr, each of the signal rate predictions from \wp\ and \zp\ production is fixed to the ratio predicted by the benchmark models. To evaluate 2D constraints on coupling strengths, the signal yields are parameterized with a set of coupling parameters ($\vec{g}$) which allow the relative contributions of each signal to vary independently.   Thus, in the 2D limit calculation, Eq.~(\ref{eqn:LLRnom}) is modified to allow the set of coupling parameters to be considered independently:
 
\begin{equation*}
\mathcal{T}^\prime = -2 \ln \left\{ L(\vec{g},\hat{\hat{\bm{\theta}}}(\vec{g})) \over L(\hat{\vec{g}},\hat{\bm{\theta}}(\hat{\vec{g}}))\right\}.
\end{equation*}
Multiple 2D coupling spaces are considered as discussed in Section~\ref{subsec:hvtcoupling}.
 
The 95\% CL upper limit contours in each coupling space are determined from $\mathcal{T}^\prime$ by normalizing signal rates to the \sigbr\ predictions of the HVT model for the specified values of $\vec{g}$ at a given point in the coupling space and calculating the value of CL$_\text{s}$ for that point.  Upper limits on coupling parameters are thus defined by contours of constant CL$_\text{s}$ = 0.05 in each coupling space considered.

% End of text imported from the .//sections/8-statistics.tex input file

% The next lines are included from the .//sections/9-results.tex input file
\section{Results}
 
\subsection{Search phase}
 
One of the advantages of combining individual search channels is that small local
excesses might reinforce each other, or larger excesses might not be corroborated.
An HVT signal scan is performed over the range of pole masses to assess the local
$p$-value for the individual channels, subcombinations, and subsequent full combination,
presented in Figure~\ref{fig:DY_pValue_Comb2}.
This stepped approach is designed to allow comparisons within the various subcombinations before the full combination, in case the new phenomena are phobic to a particular type of coupling. While a number of 1$\sigma$ and 2$\sigma$ fluctuations are observed in the individual search channels, the subcombination and full combination scans generally show fewer excesses because most of the individual excesses do not reinforce each other. For the VBF channels, which test the HVT model C, a modest increase in significance is observed for the combination at a pole mass of 1.5~\TeV with a local significance of 2.6$\sigma$, driven by the $qq \nu\nu$, $\ell\nu qq$, and $qq \ell\ell$ channels. A small excess was also observed in the previous ATLAS result~\cite{EXOT-2017-31} at a pole mass of 1.6~\TeV with a local significance of 2.9$\sigma$.
 
\begin{figure}[tb]
\centering
\subfloat[]{\includegraphics[width=0.49\textwidth]{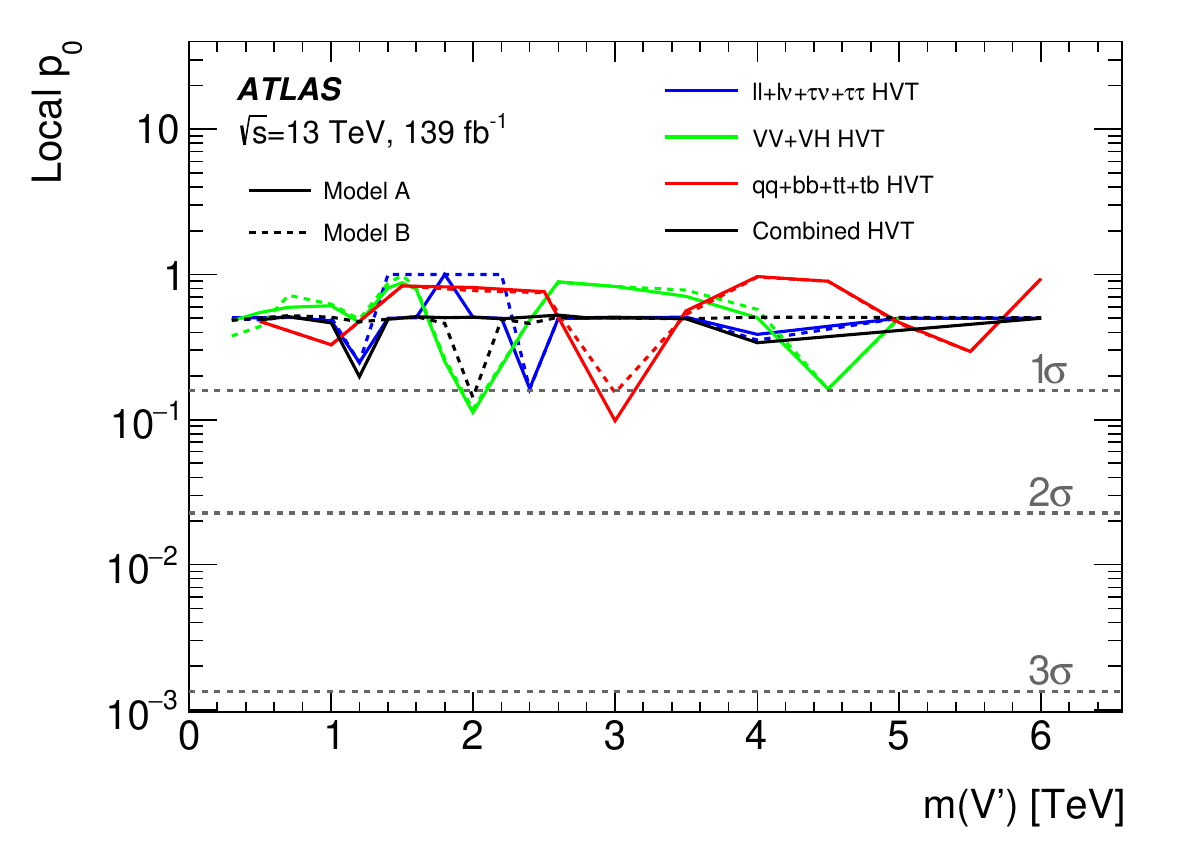}}
\subfloat[]{\includegraphics[width=0.49\textwidth]{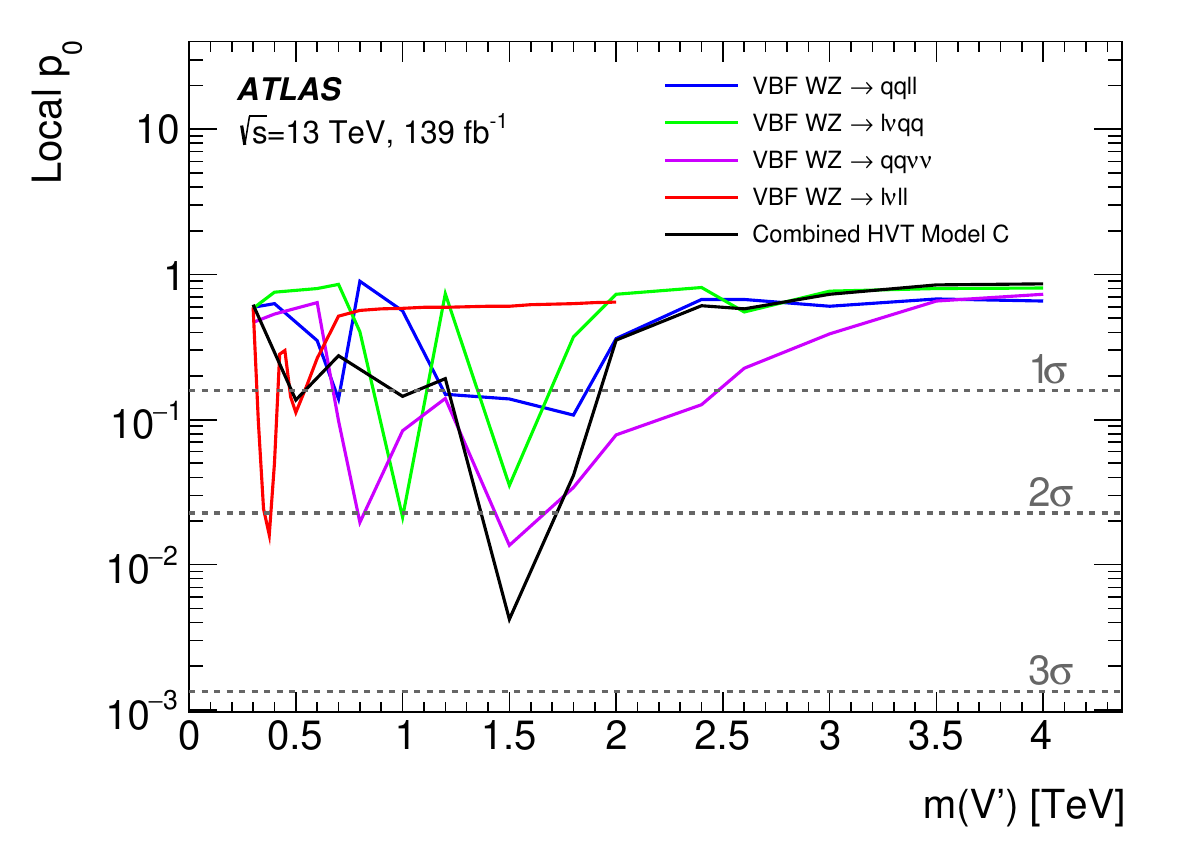}}
\caption{$p$-value scan over HVT pole masses for the subcombinations and full combination. Scans are shown for (a) the HVT model A and HVT model B, in which the scan for the full combination is slightly coarser than the individual searches due to alignment of the tested pole masses, and (b) the HVT model C.}
\label{fig:DY_pValue_Comb2}
\end{figure}
 
\subsection{1D exclusion limits}
 
As no significant excess is observed, exclusion limits are set on the signal cross-section times branching ratio for the constituent analyses of the combination. Figure~\ref{fig:DY_1D} shows the results for the $q\bar{q}$ production mode, assuming the weakly coupled HVT model A or strongly coupled HVT model B. These limits are within the region of $\Gamma/m$ $<$ 5\% and thus the narrow width approximation applies well everywhere. As expected, the purely leptonic channels dominate for the HVT model A benchmark point, while the bosonic channels dominate for the HVT model B. The resulting mass exclusion limits are presented in Table~\ref{tab:HVT_Limits}. The observed (expected) 95\% CL lower limit on the mass of $V^{\prime}$ resonances in the HVT model A is 5.8 (5.6)~\TeV\ and the corresponding limit in the HVT model B is 4.4 (4.4)~\TeV.
 
Figure~\ref{fig:DY_1D_VBF} presents the 1D exclusion limits for the VBF production mode. The HVT model C benchmark point is assumed by default, because there only bosonic channels contribute to the production and decay. As well as HVT model C ($g_H = 1$), two other points are shown to explore the bosonic coupling strength with $g_H = 3$ and $g_H = 6$. The valid range of theory cross-sections is displayed; the model starts to break down for low masses at $g_H = 1$ and $g_H = 6$. In cases where the observed limits do not quite cross the theory cross-section curve, the mass limit is extrapolated. The observed and expected 95\% CL lower limits on the mass of $V'$ resonances in the HVT model C are 0.4~\tev\ and 0.5~\tev, respectively. For $g_H = 3$ the observed and expected 95\% CL lower limits are 1.0~\tev\ and 1.1~\tev, respectively, while for $g_H = 6$ they are 1.5~\tev\ and 1.8~\tev, respectively.
 
\begin{figure}[tb]
\centering
\subfloat[]{\includegraphics[width=0.49\textwidth]{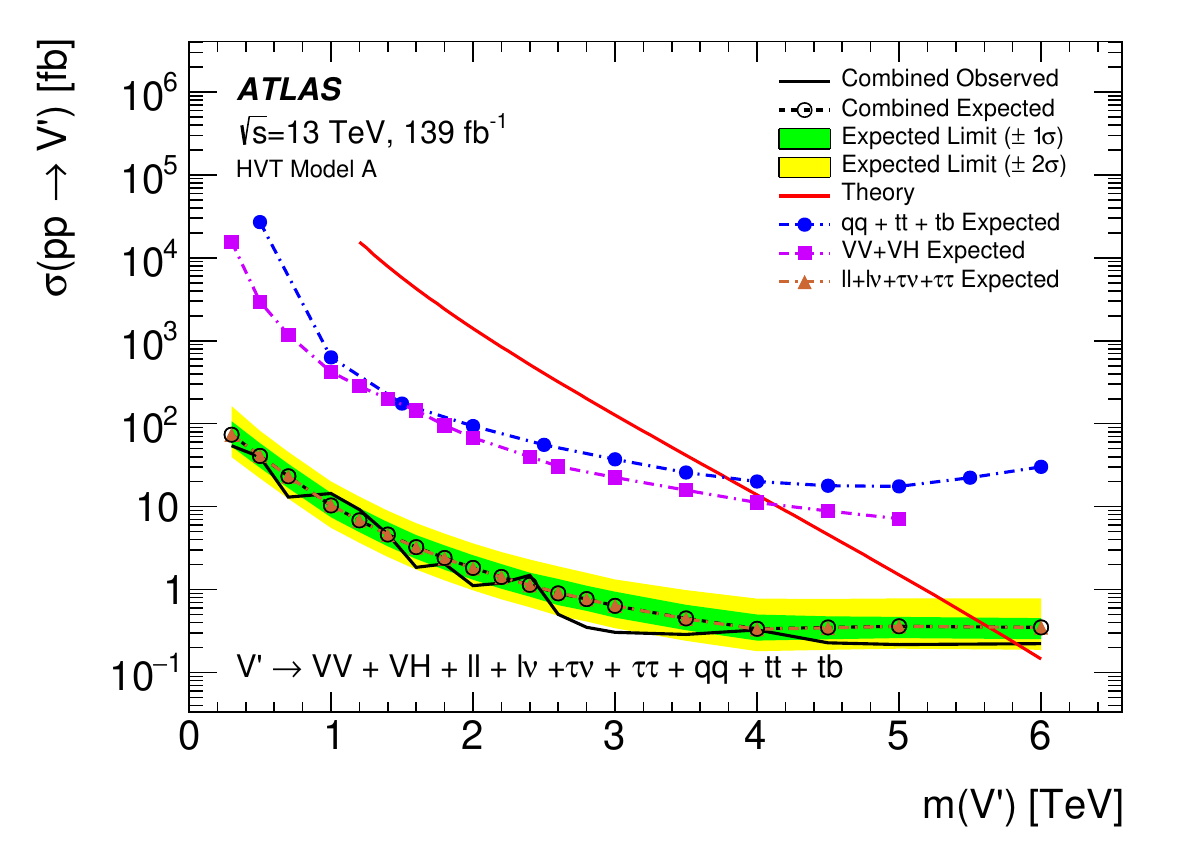}}
\subfloat[]{\includegraphics[width=0.49\textwidth]{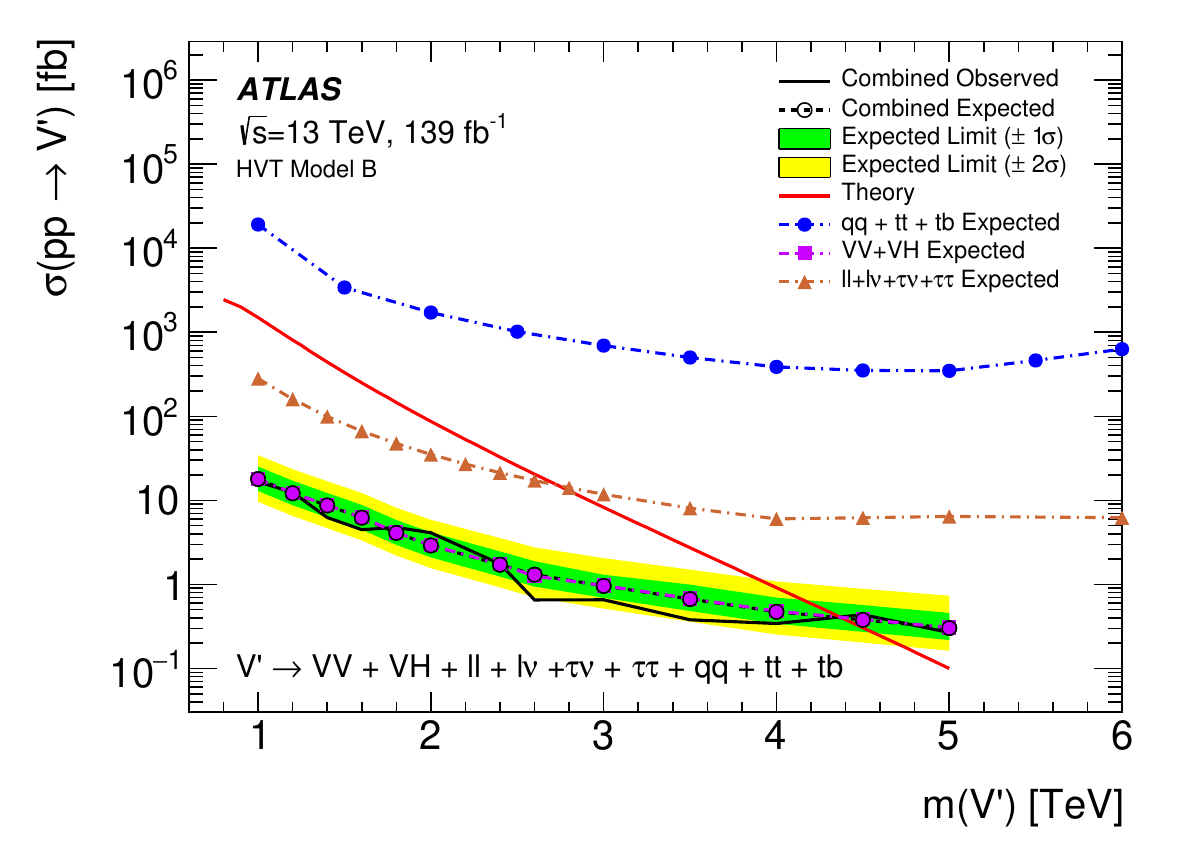}}
\caption{Observed and expected 95\% CL upper limits on the $V'$ cross-section versus pole mass
for the subcombinations and full combination, assuming the (a) HVT model A, or (b) HVT model B benchmark points, and $q\bar{q}$ production mode.}
\label{fig:DY_1D}
\end{figure}
 
\begin{figure}[!h]
\vspace{-0.5cm}
\centering
\includegraphics[width=0.49\textwidth]{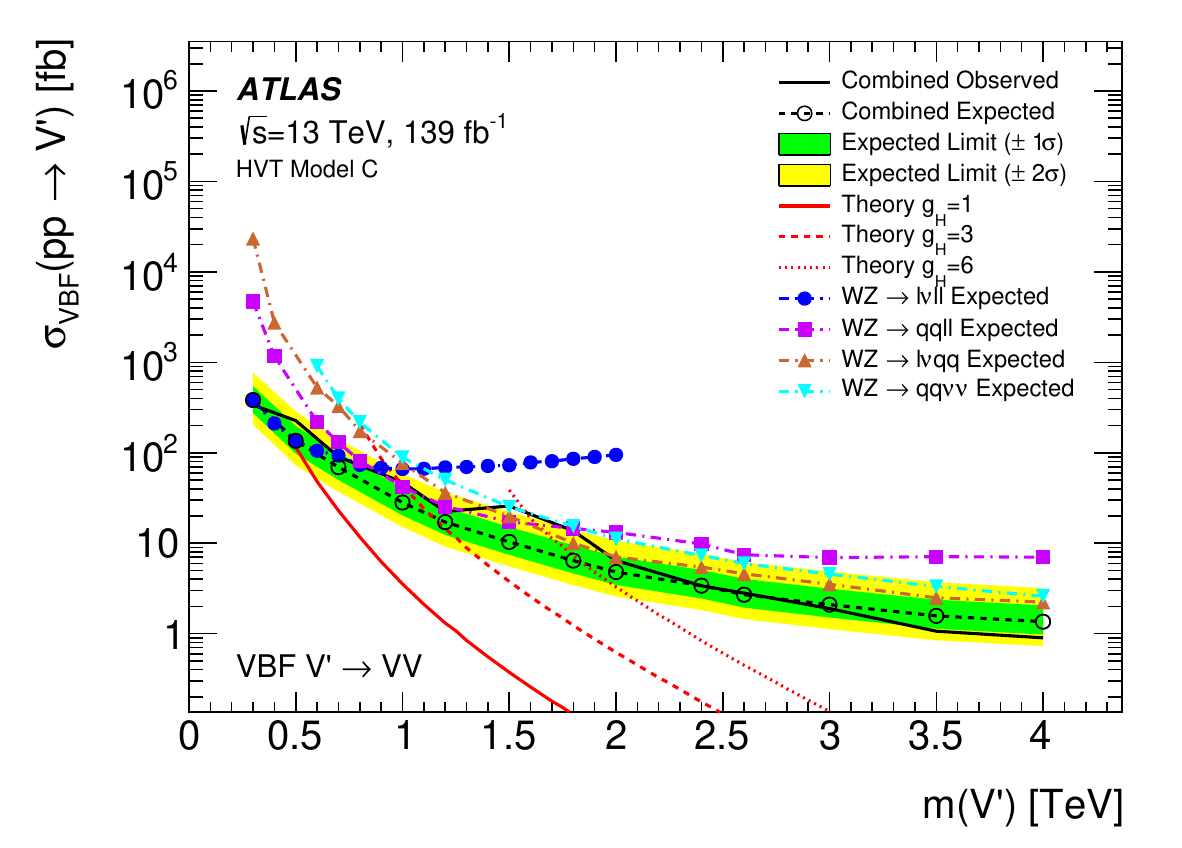}
\caption{Observed and expected 95\% CL upper limits on the $V'$ cross-section times branching fraction in
the VBF production mode for the HVT model C benchmark point.}
\label{fig:DY_1D_VBF}
\end{figure}
 
\begin{table}[h!]
\begin{center}
\caption{Observed and expected 95\% CL lower limits on the $V'$ mass at the HVT model A and HVT model B benchmark points for the $q\bar{q}$ production mode. Limits are quoted to the nearest 100~\GeV.}
\label{tab:HVT_Limits}
\begin{tabular}{ccccc}
\toprule
\multirow{2}{*}{Channel} & \multicolumn{2}{c}{HVT model A exclusion limit} & \multicolumn{2}{c}{HVT model B exclusion limit} \\
& Observed [TeV] & Expected [TeV] & Observed [TeV] & Expected [TeV] \\
\midrule
$VV$                              & 4.1       & 4.0     & 4.3     & 4.2   \\
$VH$                              & 3.6       & 3.5     & 3.9     & 3.9   \\ \hline
Bosonic                           & 4.3       & 4.1     & 4.4     & 4.4   \\
Leptonic                          & 5.8       & 5.6     & 3.2     & 2.7     \\
Quarkonic                         & 4.1       & 3.8     & -     & -     \\ \hline
Full combination                  & 5.8       & 5.6     & 4.4     & 4.4   \\
\bottomrule
\end{tabular}
\end{center}
\end{table}
 
\subsection{2D exclusion limits}
 
Exclusion limits are also presented in various 2D coupling planes for a representative signal pole mass of 3~\tev\
for the different subcombinations of channels; for the full combination, they are presented for masses ranging between 2~\tev\ and 5~\tev, depending on the sensitivity.
 
\subsubsection{Bosonic subcombination}
 
\begin{figure}[bht!]
\vspace{-0.5cm}
\centering
\subfloat[]{\includegraphics[width=0.45\textwidth]{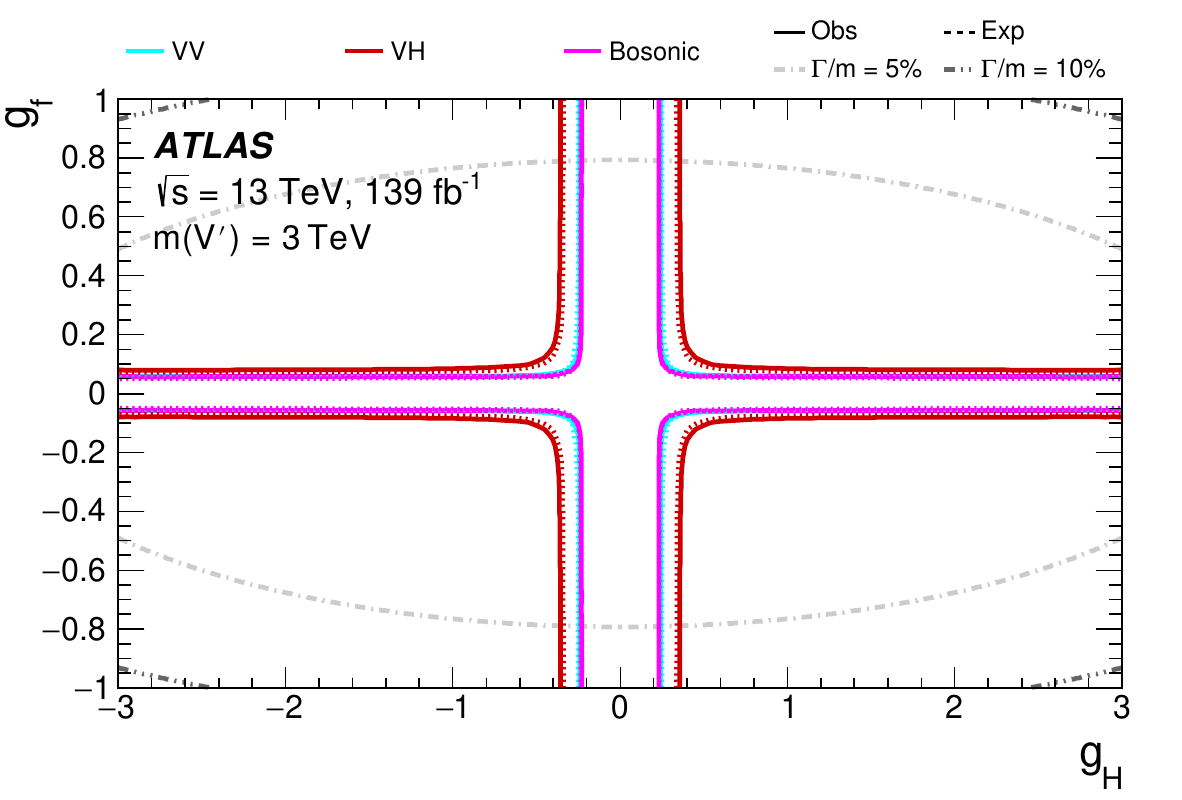}}
\subfloat[]{\includegraphics[width=0.45\textwidth]{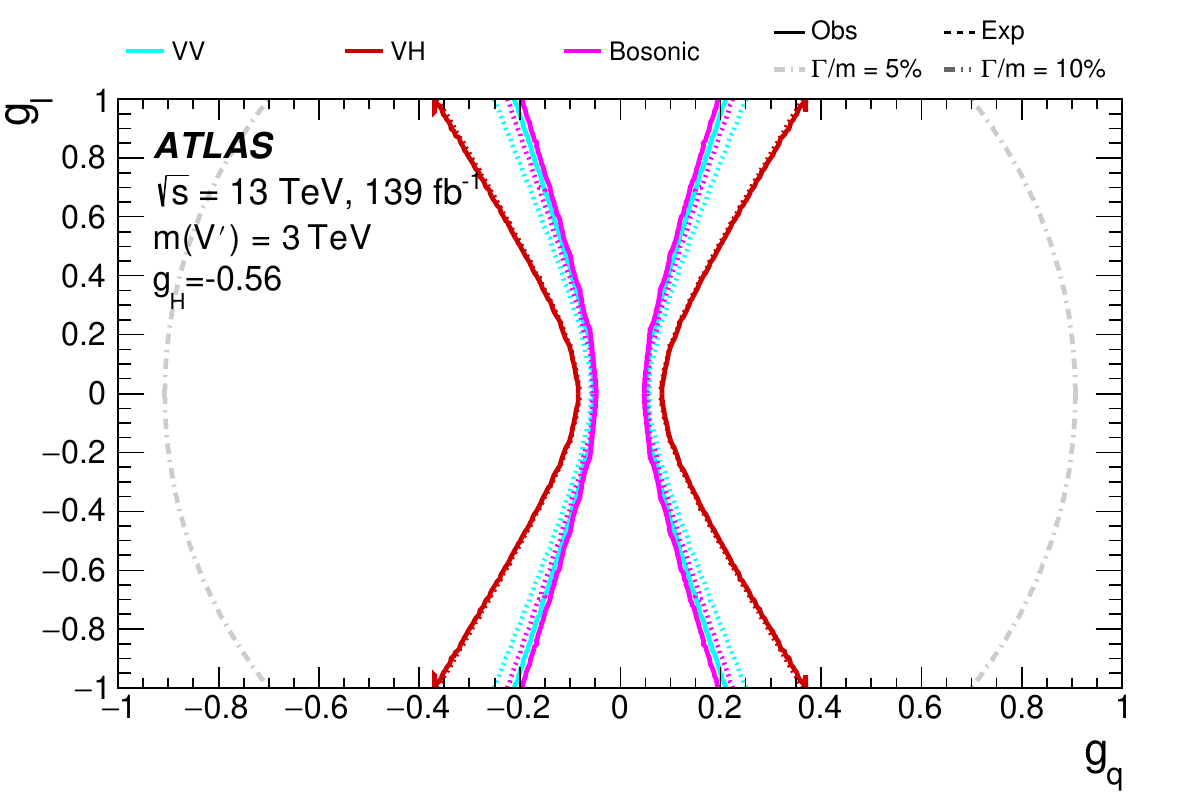}} \\
\subfloat[]{\includegraphics[width=0.45\textwidth]{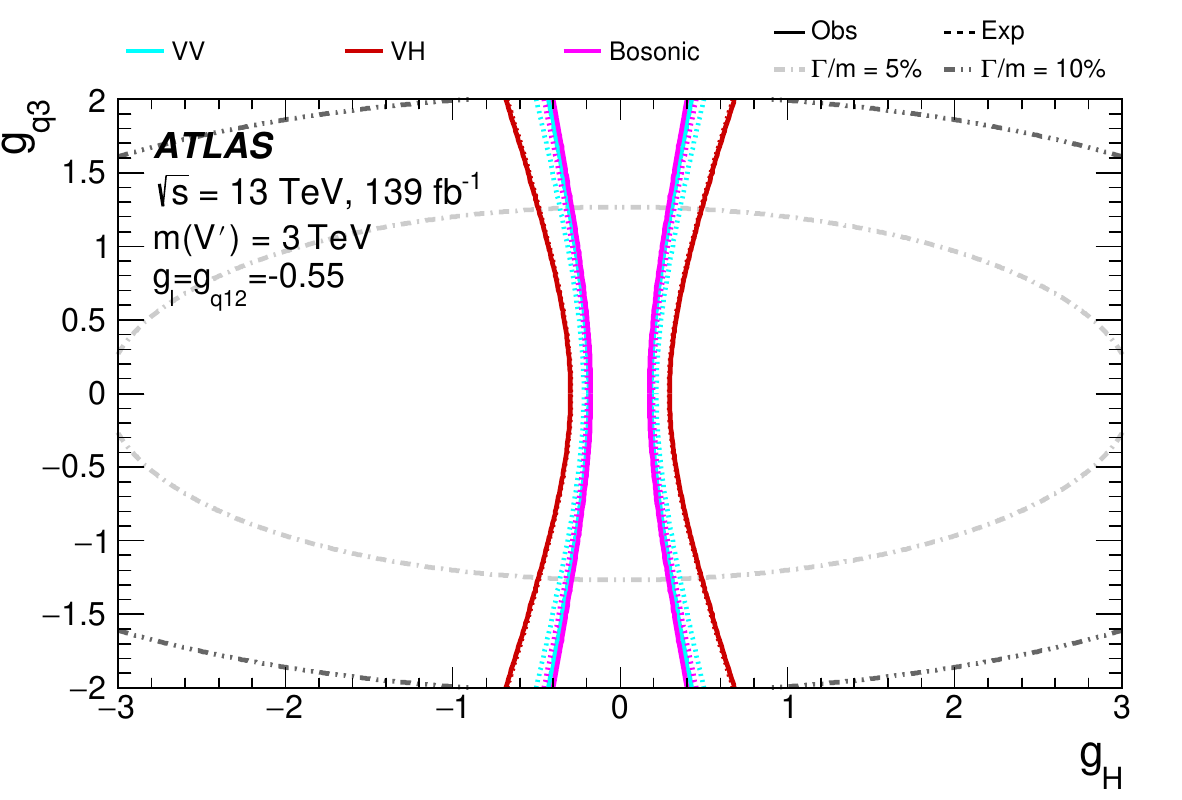}}
\caption{95\% CL observed and expected upper limit contours for the bosonic subcombination in the 2D coupling plane for
(a) $\{g_H, g_f\}$, (b) $\{g_q, g_{\ell}\}$ and (c) $\{g_H, g_{q3}\}$
at a signal pole mass of 3~\TeV\ for the $q\bar{q}$ production mode. Bosonic here refers to the combination of $VV+VH$ channels.
The dashed grey lines show the region where the resonance natural width is either 5\% or 10\% of the pole mass. Note that in (b) the 10\% width line is outside of the shown range.}
\label{fig:DY_2D_Bosonic}
\end{figure}
 
The exclusion limits from the $VV$ channels are generally stronger than those from the $VH$ channels, but the latter
lead to improvements in sensitivity when combined, as presented in Figure~\ref{fig:DY_2D_Bosonic}
for the $\{g_H, g_f\}$, $\{g_q, g_{\ell}\}$ and $\{g_{q3}, g_{\ell}\}$ coupling planes.
This improvement grows with increasing signal pole mass.
 
\clearpage
 
\subsubsection{Leptonic subcombination}
 
\begin{figure}[b]
\centering
\subfloat[]{\includegraphics[width=0.49\textwidth]{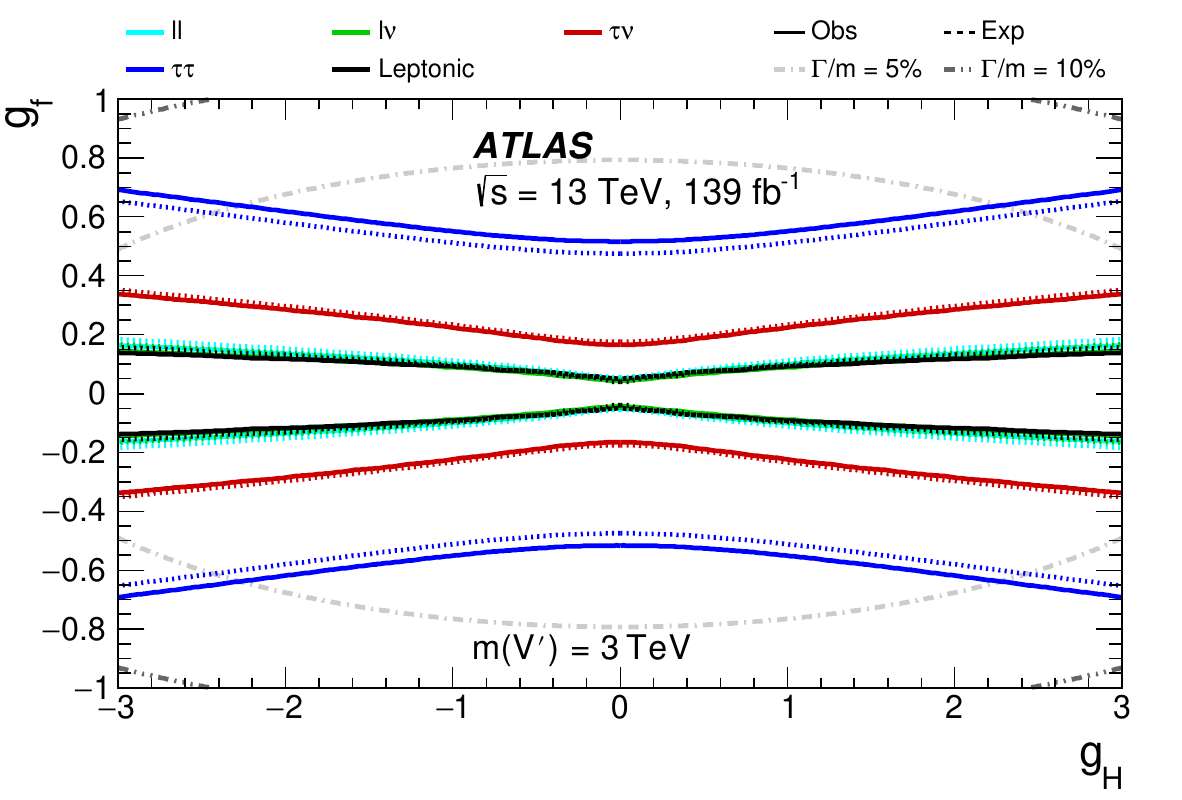}}
\subfloat[]{\includegraphics[width=0.49\textwidth]{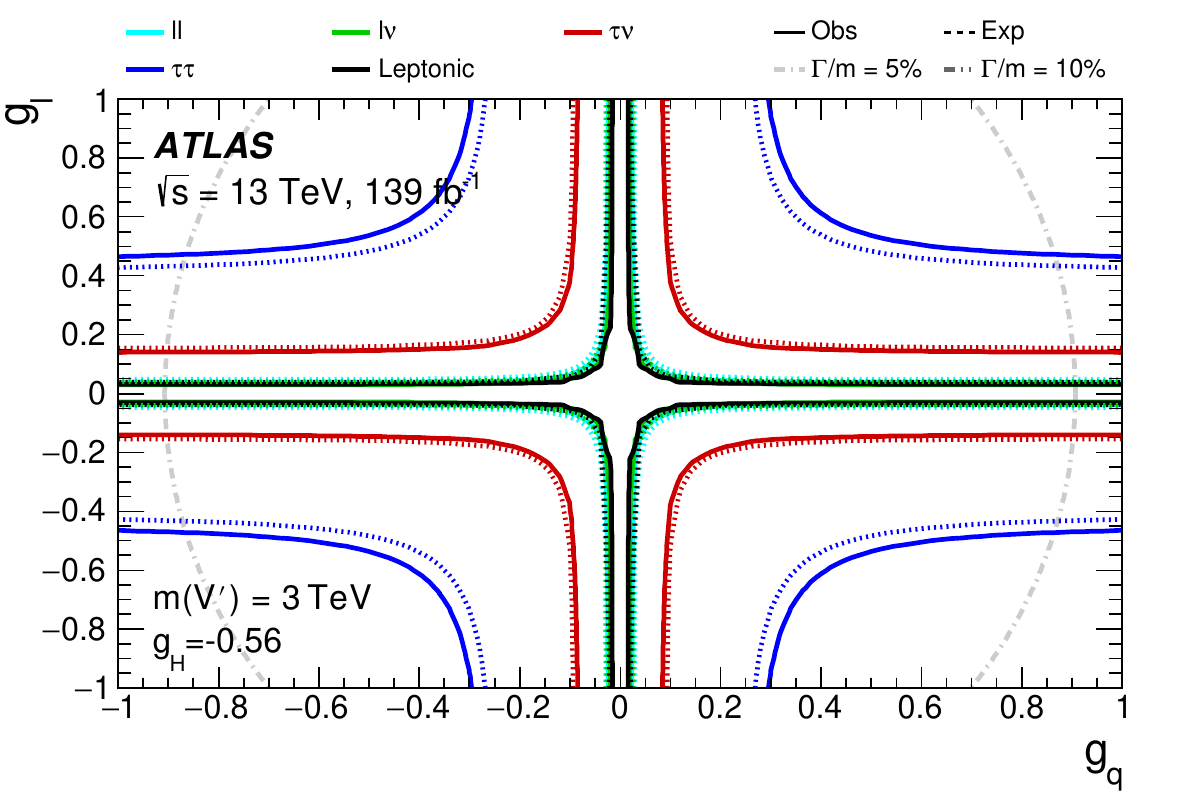}} \\
\subfloat[]{\includegraphics[width=0.49\textwidth]{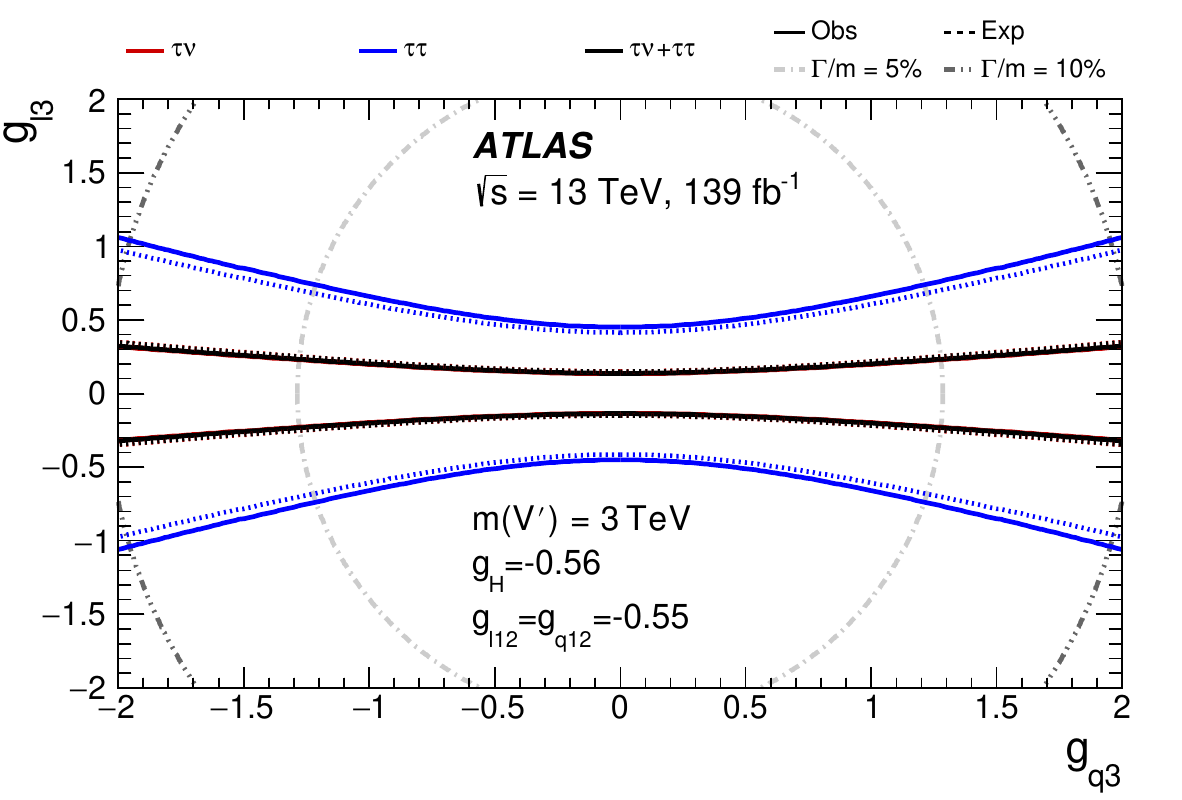}}
\caption{95\% CL observed and expected upper limit contours for the leptonic subcombination in the 2D coupling plane for
(a) $\{g_H, g_f\}$, (b) $\{g_q, g_{\ell}\}$ and (c) $\{g_{q3}, g_{\ell 3}\}$
at a signal pole mass of 3\,\TeV\
for the $q\bar{q}$ production mode. Leptonic here refers to the combination of $\ell\ell+\ell\nu+\tau\tau+\tau\nu$ channels. The dashed grey lines show the region where the resonance natural width is either 5\% or 10\% of the pole mass. Note that in (b) the 10\% width line is outside of the shown range.
In (c) the limit contours from $\tau\nu$ and the $\tau\nu+\tau\tau$ combination overlap exactly, making the former not visible.}
\label{fig:DY_2D_Leptonic}
\end{figure}
 
The leptonic subcombination is presented in Figure~\ref{fig:DY_2D_Leptonic}
for the $\{g_H, g_f\}$, $\{g_q, g_{\ell}\}$ and $\{g_{q3}, g_{\ell 3}\}$ coupling planes.
The newly included $\tau\nu$ and $\tau\tau$ channels are weaker than other channels in the coupling planes
that consider all fermions, but become prominent in the third-generation-specific coupling planes.
The overall subcombination is dominated by the charged-current channels because of the higher
expected $W'$ production cross-sections. The addition of the neutral-current channels leads to improvements of up to 15\%--20\% in the constraints,
depending on the coupling plane.
For most pole masses tested in this paper, the leptonic channels exhibit a slight deficit,
which causes the observed limits to be stronger than expected.
 
\subsubsection{Quarkonic subcombination}
 
The quarkonic subcombination is performed for the first time in this paper.
It is generally dominated by the $tb$ channels,
especially at higher masses, mainly due to $W'$ production having a
higher cross-section than $Z'$ production. However, the
other quarkonic channels provide complementary sensitivity, leading to improved
constraints from the combination of all channels.
Figure~\ref{fig:DY_2D_Quarkonic} presents the results in the five coupling planes
considered in this publication. The constraints on the $\{g_H, g_f\}$, $\{g_q, g_{\ell}\}$, $\{g_H, g_{q3}\}$
and $\{g_{q3}, g_{\ell 3}\}$ couplings planes complement those obtained in the bosonic and leptonic channel subcombinations.
The exclusion limit contours in the fifth coupling plane, $\{g_{q12}, g_{q3}\}$, are constrained by quarkonic channels only.
This coupling plane is particularly interesting because the $qq$ channel is the only one
sensitive in the region where the third-generation coupling vanishes, while
the third-generation channels naturally provide a tighter constraint in the region
where $g_{q3}$ is not small. The subcombination of all quarkonic channels thus
results in a narrow vertical exclusion band in Figure~\ref{fig:DY_2D_Quarkonic}, highlighting the excellent complementarity
and providing a large increase in sensitivity in comparison with the separate channels.
 
\begin{figure}[hbt!]
\centering
\subfloat[]{\includegraphics[width=0.49\textwidth]{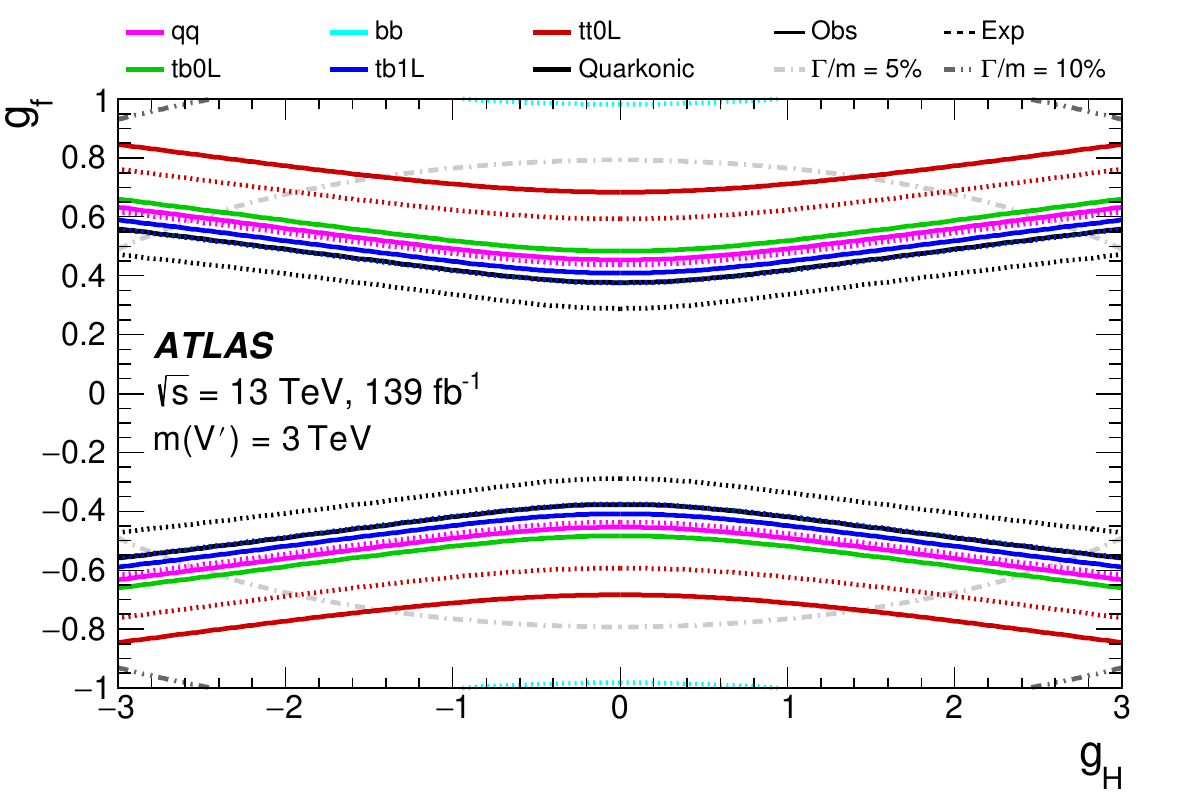}}
\subfloat[]{\includegraphics[width=0.49\textwidth]{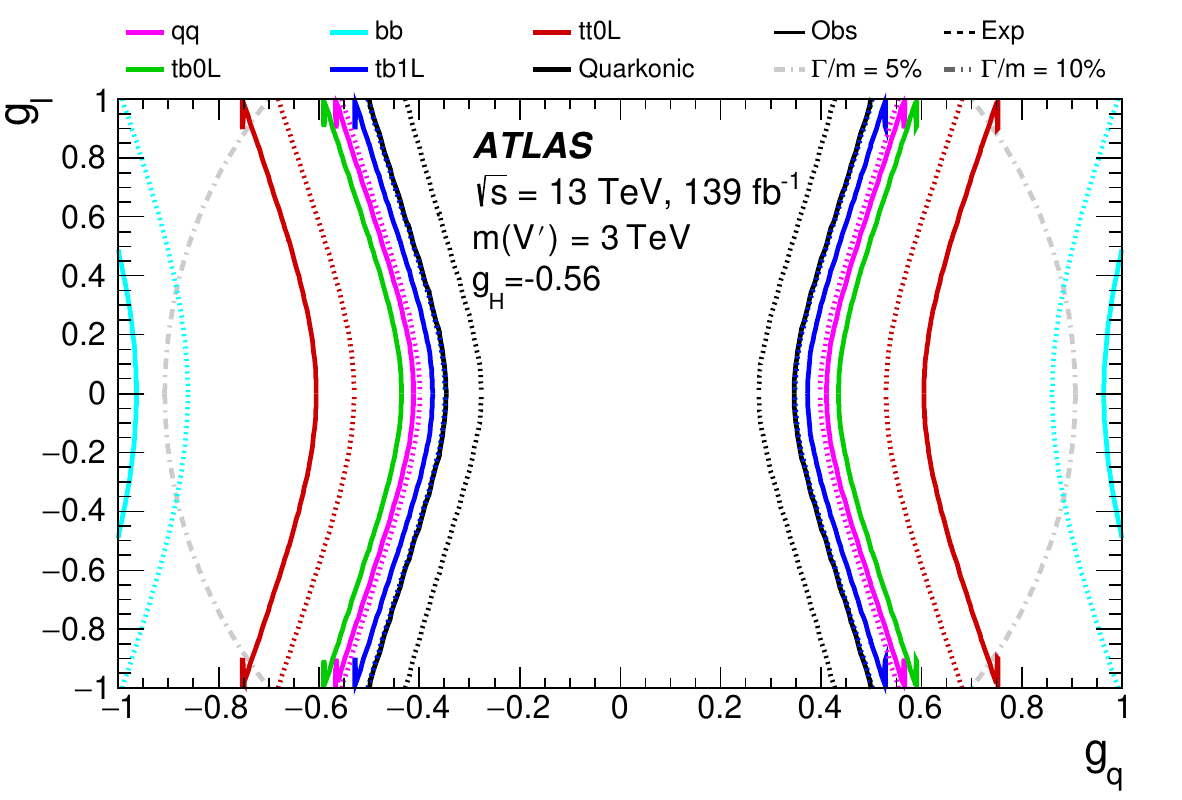}} \\
\subfloat[]{\includegraphics[width=0.49\textwidth]{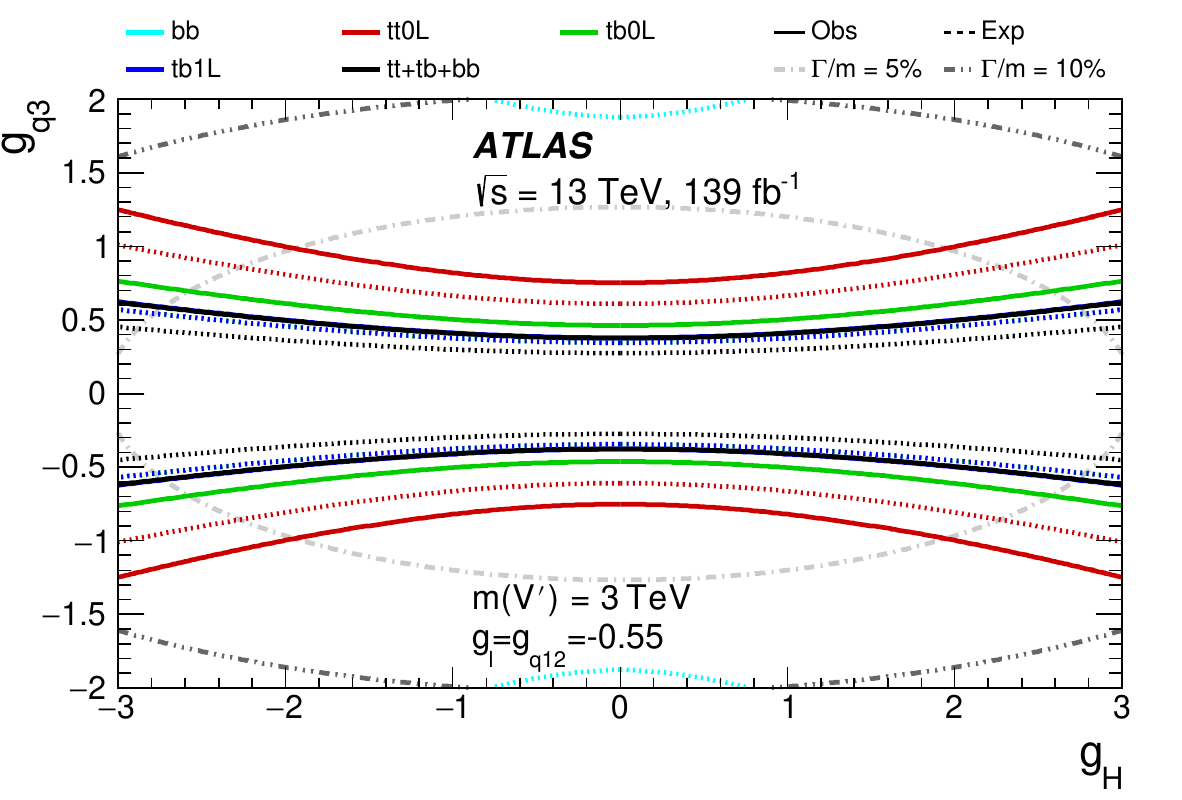}}
\subfloat[]{\includegraphics[width=0.49\textwidth]{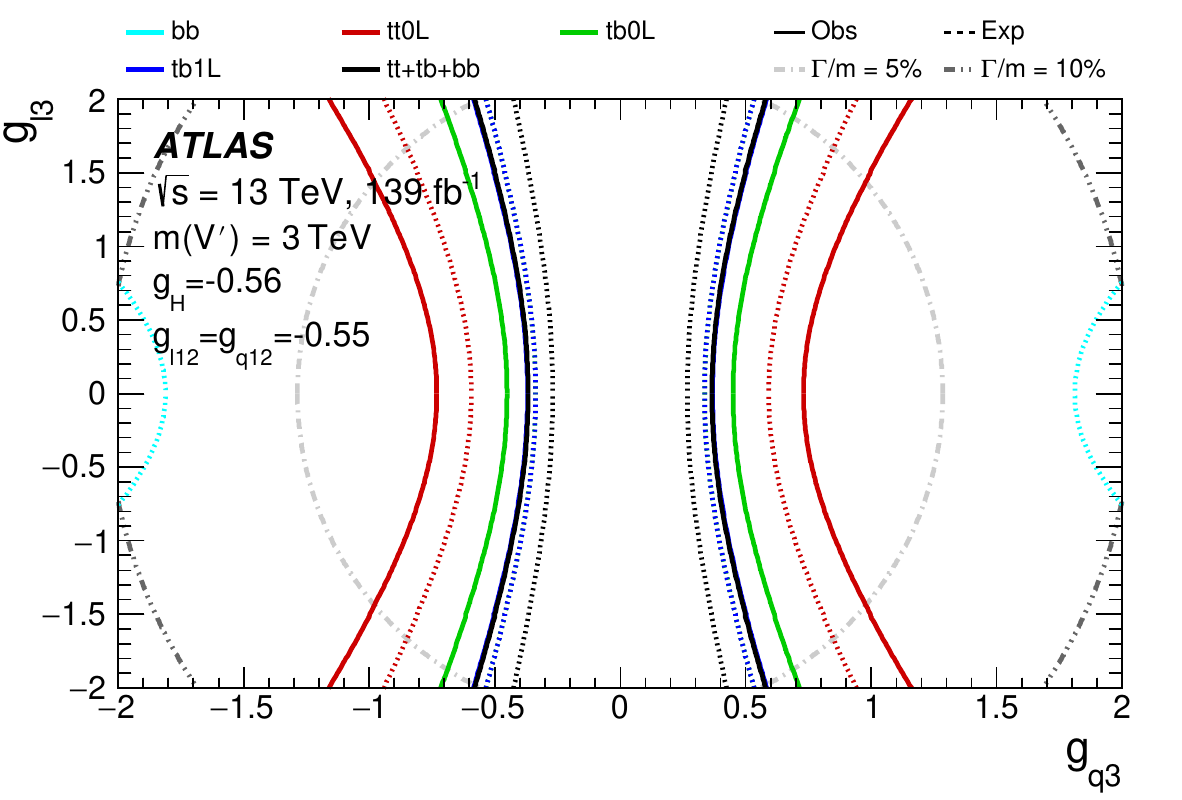}}\\
\subfloat[]{\includegraphics[width=0.49\textwidth]{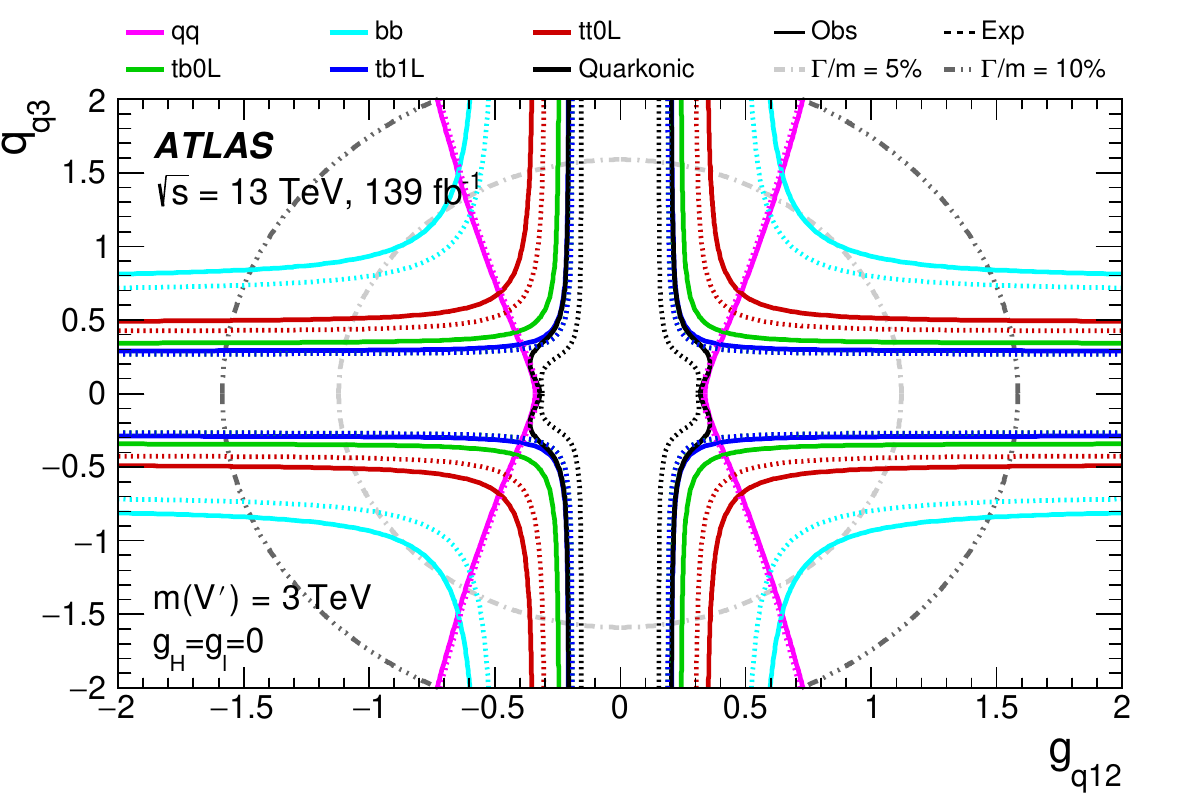}}
\caption{95\% CL observed and expected upper limit contours for the quarkonic subcombination in the 2D coupling plane for
(a) $\{g_H, g_f\}$, (b) $\{g_q, g_{\ell}\}$, (c) $\{g_H, g_{q3}\}$
(d) $\{g_{q3}, g_{\ell 3}\}$ and (e) $\{g_{q12}, g_{q3}\}$
at a signal pole mass of 3~\TeV\
for the $q\bar{q}$ production mode. Quarkonic here refers to the combination of $qq+bb+tt+tb$ channels.
The dashed grey lines show the region where the resonance natural width is either 5\% or 10\% of the pole mass.}
\label{fig:DY_2D_Quarkonic}
\end{figure}
 
\subsubsection{Full combination}
 
The full combination of bosonic, leptonic and quarkonic searches is presented for the first time in this paper. The complementarity of various channels in the combination is described below for the following four coupling planes:
$\{g_H, g_f\}$, $\{g_q, g_{\ell}\}$, $\{g_H, g_{q3}\}$, and $\{g_{q3}, g_{\ell 3}\}$.
 
\begin{figure}[hbt!]
\centering
\subfloat[]{\includegraphics[width=0.49\textwidth]{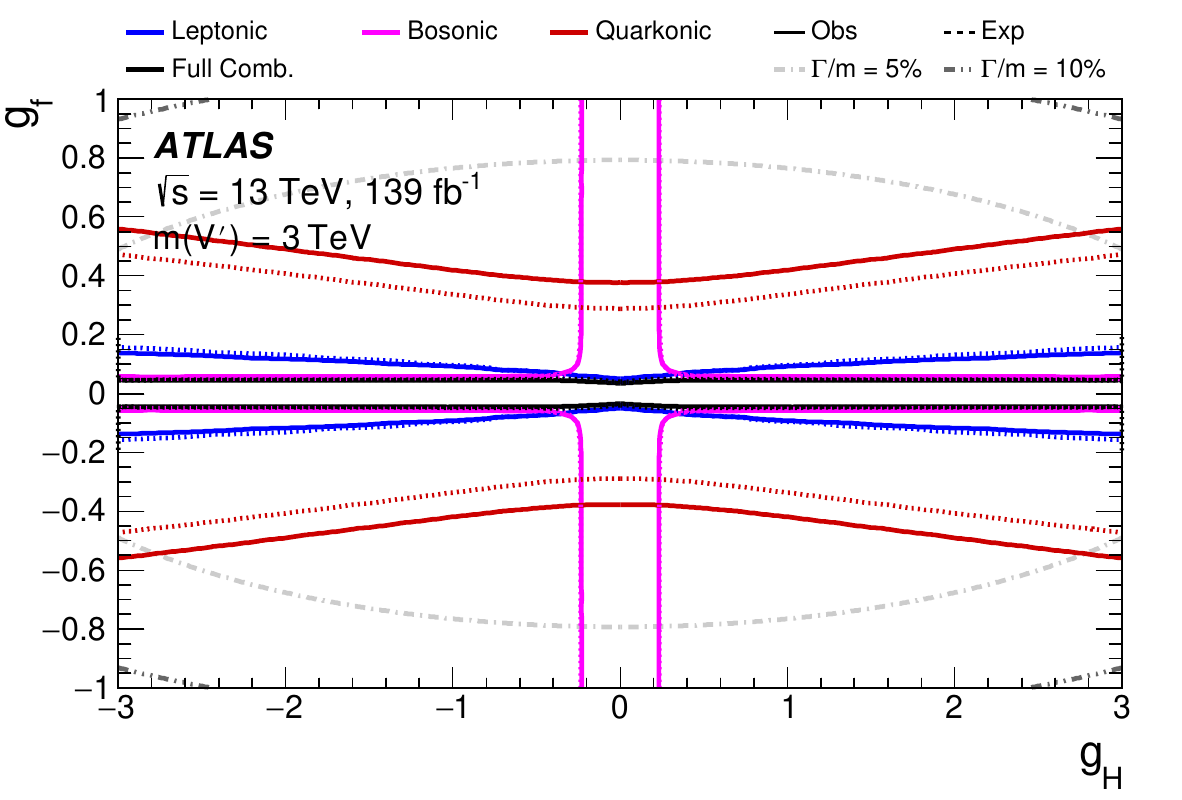}}
\subfloat[]{\includegraphics[width=0.49\textwidth]{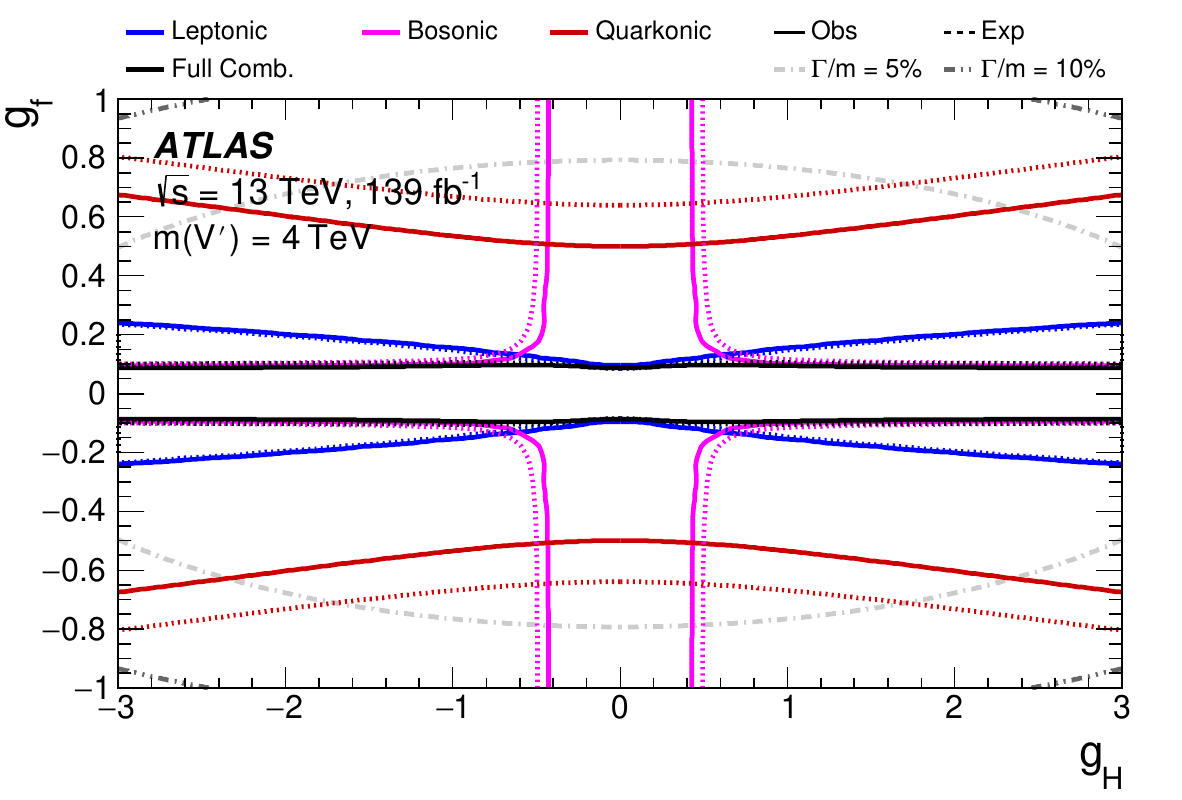}} \\
\subfloat[]{\includegraphics[width=0.49\textwidth]{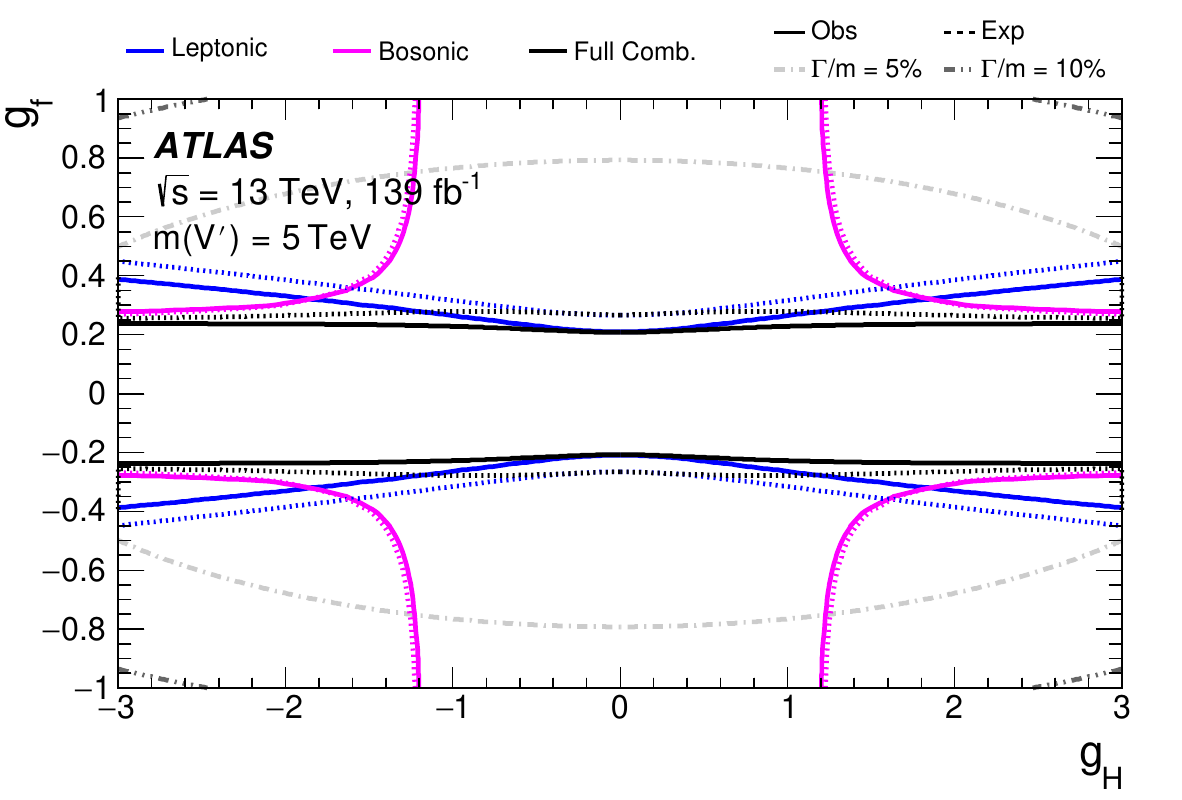}}
\caption{95\% CL observed and expected upper limit contours for the full combination in the 2D coupling plane for $\{g_H, g_f\}$ at a signal pole mass of (a) 3~\TeV, (b) 4~\TeV, and (c) 5~\TeV, for the $q\bar{q}$ production mode. Full combination here refers to the combination of leptonic, bosonic, and quarkonic channels. The dashed grey lines show the region where the resonance natural width is either 5\% or 10\% of the pole mass.}
\label{fig:DY_2D_gHgF_Comb}
\end{figure}
 
Figure~\ref{fig:DY_2D_gHgF_Comb} shows the full combination results for the $\{g_H, g_f\}$ plane for
pole masses of 3~\TeV, 4~\TeV, and 5~\TeV. These results highlight the complementarity of
the bosonic channels and the other channels, as the bosonic channels can constrain the couplings more tightly
in the region where $g_f$ tends to zero, while the leptonic and quarkonic channels remain sensitive when
$g_H$ vanishes. The resulting combination of all channels leads to only very narrow horizontal bands
which cannot be excluded given that setting $g_f = 0$ would also extinguish the $q\bar{q}$
production mechanism. Only VBF channels can contribute in the low $g_f$ region, but their
sensitivity is currently too small to appear in the coupling-parameter space shown in these plots.
For a signal pole mass of 3~\TeV, the constraints are so tight that the combination does not
give much improvement on the individual channels. At high
masses, such as 5~\TeV, the combination can improve the limit by up to 40\%.
This result significantly improves on the earlier $VV+VH+\ell\ell+\ell\nu$
combination~\cite{EXOT-2017-31},
which used 36~\ifb\ of $\sqrt{s} = 13$~\TeV\ data. For signal pole masses of 3~\TeV\ and 4~\TeV,
the expected limit on $g_f$ is improved by 10\%--50\% depending on the value of $g_H$,
while for a pole mass of 5~\TeV\ the limit is improved by 40\%--60\%.
The main increase in sensitivity comes from the fermionic channels, which have
now reached a level of sensitivity such that they effectively \enquote{pinch in} the limit contours near $g_H = 0$.
While there are modest increases in sensitivity for intermediate pole masses,
which is understandable given that the limits are already very stringent in this region, the biggest improvement
is obtained at high mass, where the combination of channels improves the sensitivity in the regions
where no individual channel is dominant.
 
\begin{figure}[b]
\centering
\subfloat[]{\includegraphics[width=0.49\textwidth]{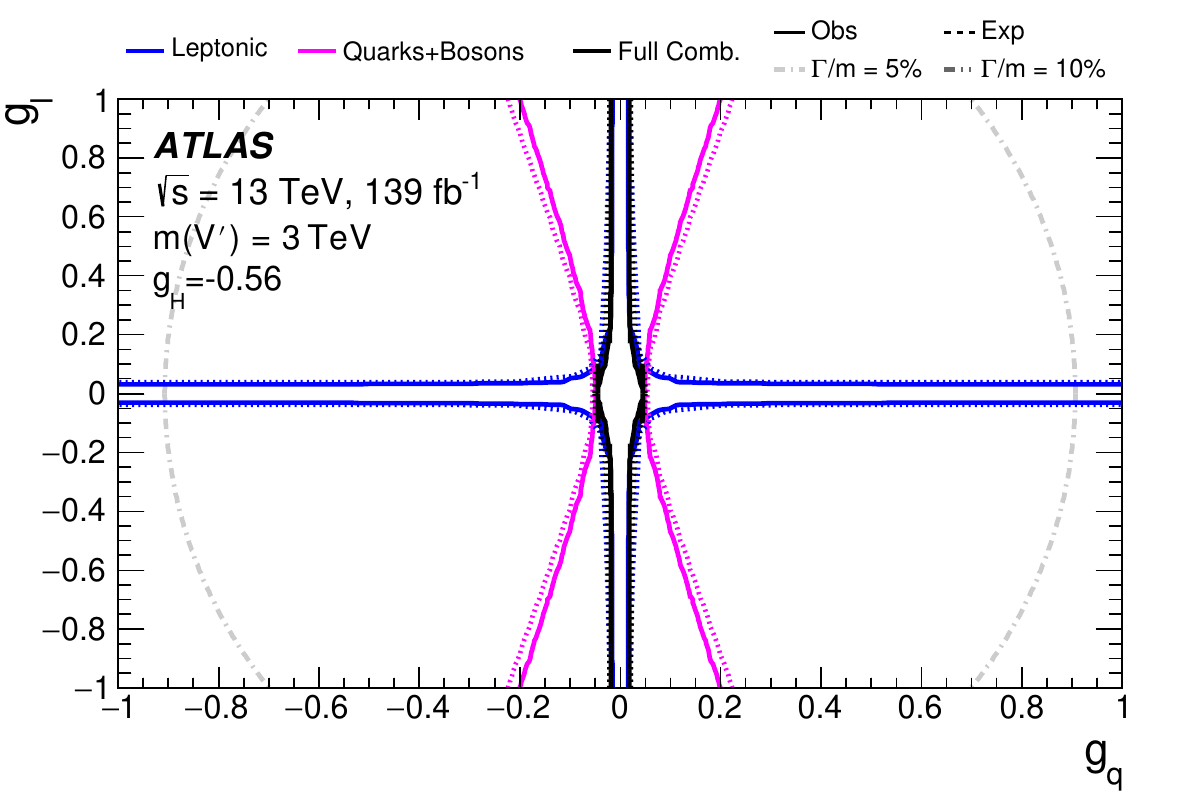}}
\subfloat[]{\includegraphics[width=0.49\textwidth]{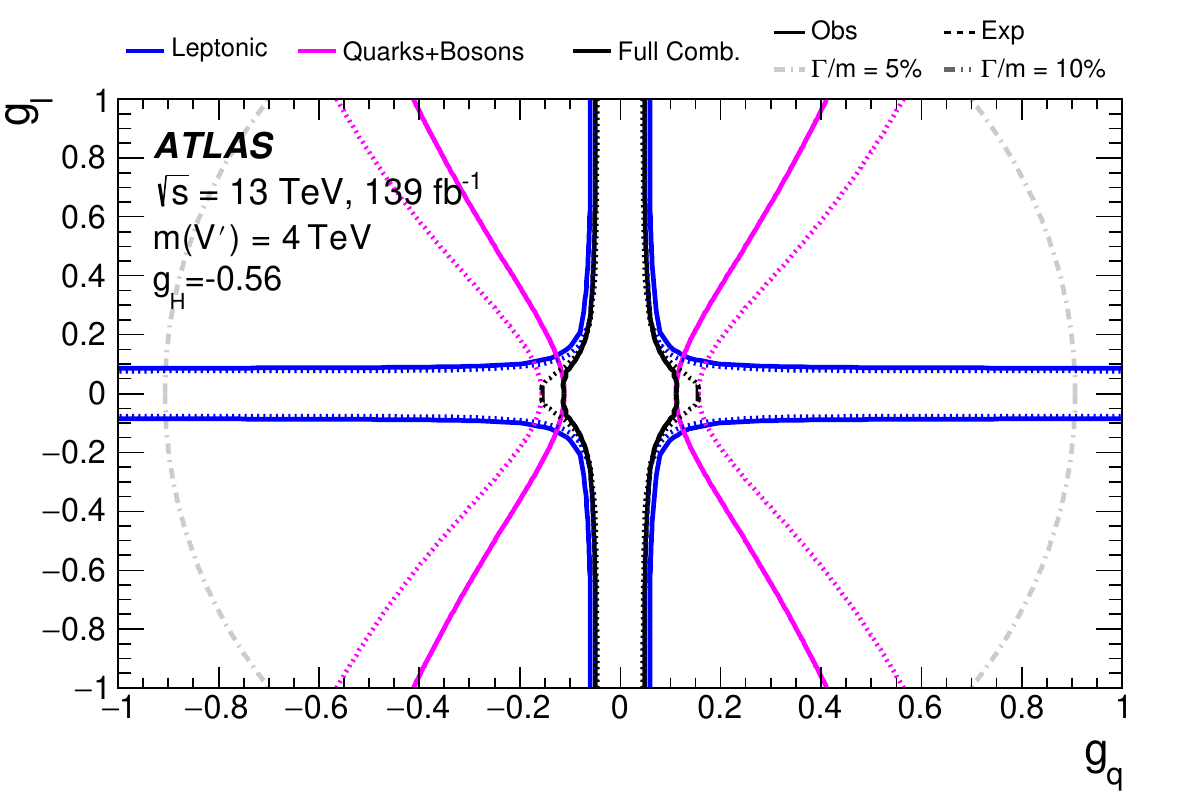}}
\caption{95\% CL observed and expected upper limit contours for the full combination
in the 2D coupling plane for $\{g_q, g_{\ell}\}$ at a signal pole mass of (a) 3~\TeV, and (b) 4~\TeV,
for the $q\bar{q}$ production mode. Full combination here refers to the combination of leptonic, bosonic, and quarkonic channels. The dashed grey line shows the region where the resonance natural
width is 5\% of the pole mass.}
\label{fig:DY_2D_gQgL_Comb}
\end{figure}
 
Figure~\ref{fig:DY_2D_gQgL_Comb} shows the full combination results for the $\{g_q, g_\ell\}$ plane for
pole masses of 3~\TeV\ and 4~\TeV. The 2~\TeV pole mass is fully excluded for this point in parameter space,
while only the leptonic channels enter in the 5~\TeV\ plane, which
renders the combination unnecessary. The mass points shown highlight the
complementarity of the channels in the full combination: the bosonic and quarkonic channels
set strong constraints in the limit where $g_{\ell}$ tends to zero, while the leptonic channels
remain dominant for smaller values of $g_q$, provided $g_{\ell}$ is not small. The bosonic limits are observed to
sweep outwards in $g_q$ with increasing pole mass, as the cross-section in these regions falls rapidly.
 
\begin{figure}[b]
\centering
\subfloat[]{\includegraphics[width=0.49\textwidth]{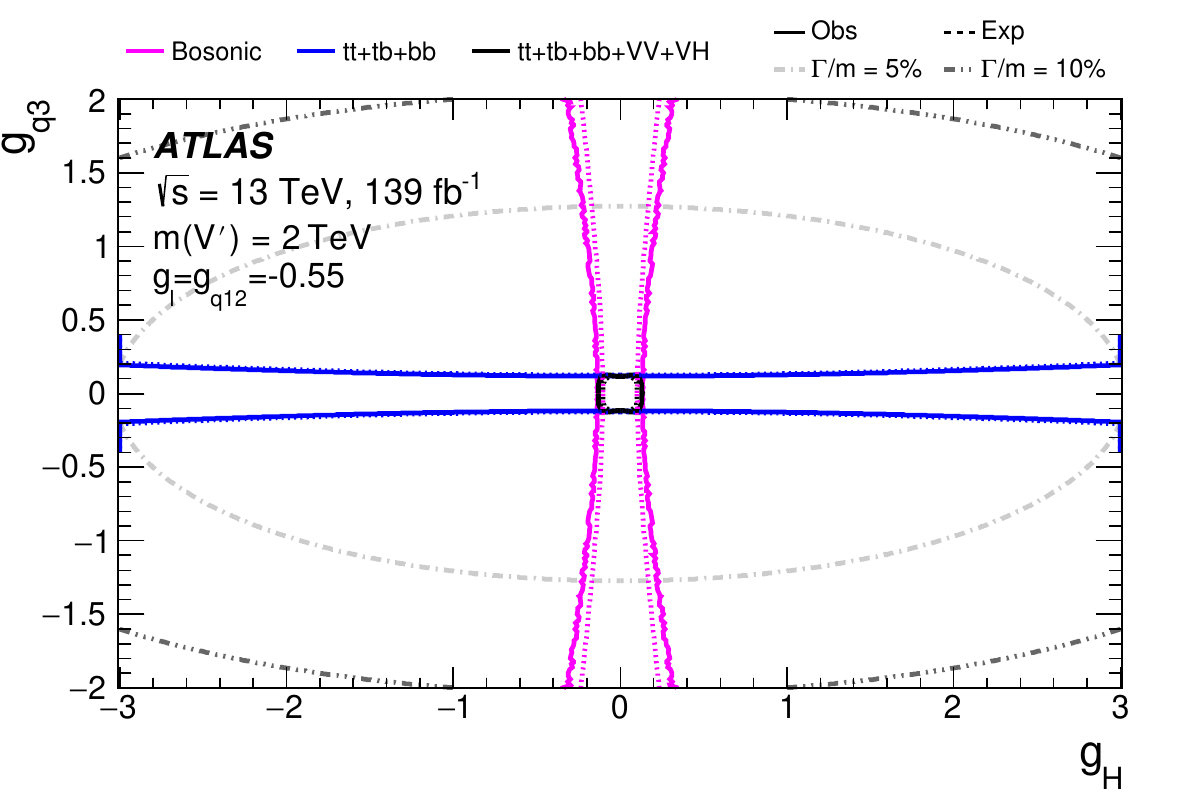}}
\subfloat[]{\includegraphics[width=0.49\textwidth]{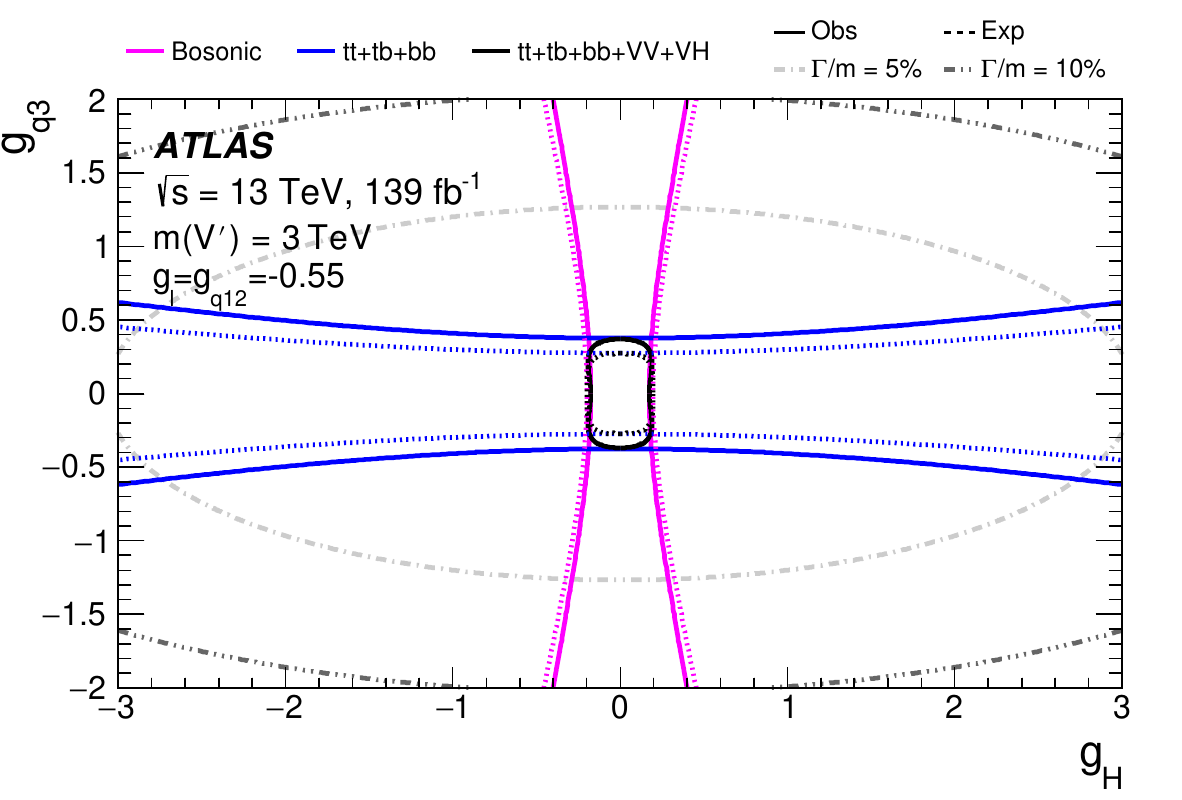}}\\
\subfloat[]{\includegraphics[width=0.49\textwidth]{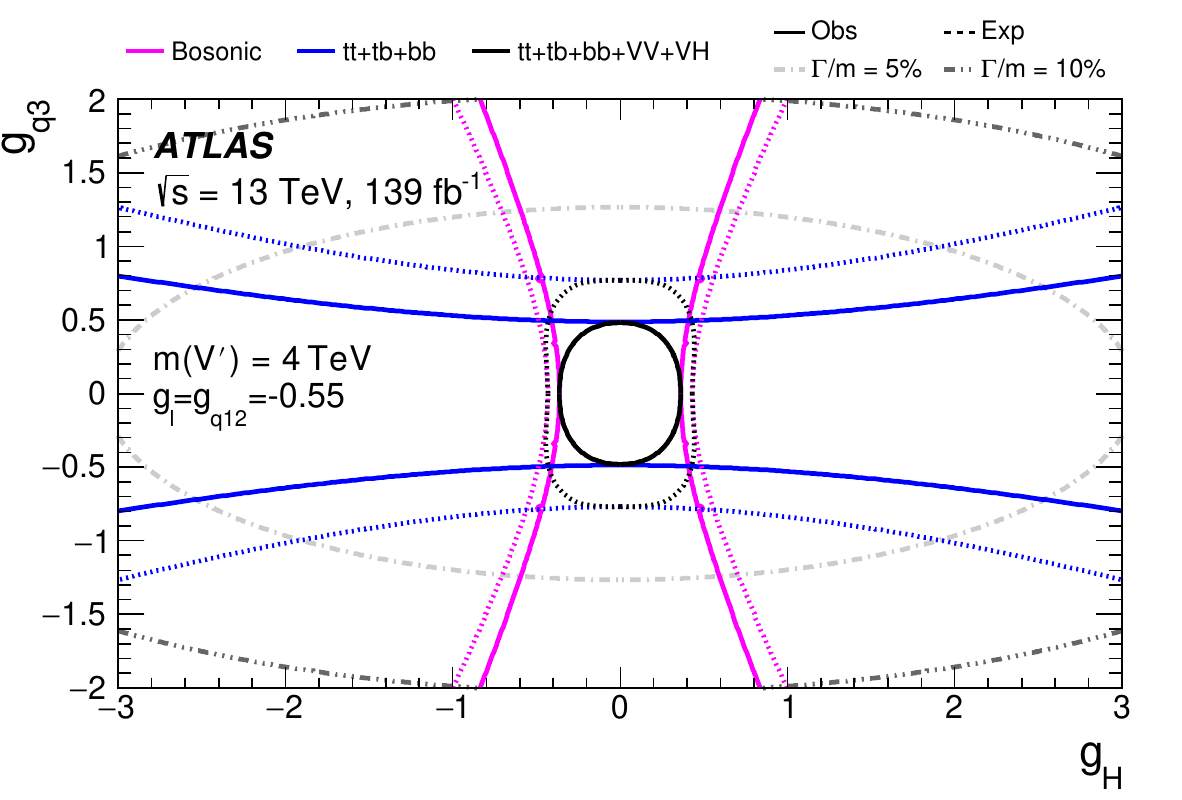}}
\caption{95\% CL observed and expected upper limit contours for the full combination in the
2D coupling plane for $\{g_H, g_{q3}\}$ at a signal pole mass of (a) 2~\TeV, (b) 3~\TeV, and
(c) 4~\TeV\ for the $q\bar{q}$ production mode. Bosonic here refers to the combination of $VV+VH$ channels. The dashed grey lines show the region where
the resonance natural width is either 5\% or 10\% of the pole mass.}
\label{fig:DY_2D_gHgQ3_Comb}
\end{figure}
 
The $\{g_H, g_{q_3}\}$ plane, shown in Figure~\ref{fig:DY_2D_gHgQ3_Comb},
is a particularly nice example of how the separate channels reinforce each other in the full combination.
The bosonic channels constrain all but a vertical strip where $g_H$ tends to zero, while
the third-generation quark channels constrain all but a horizontal strip where $g_{q_3}$ tends to zero.
These subcombinations provide unrivalled sensitivity in their respective regions of phase space, and it is only when
both $g_H$ and $g_{q_3}$ become smaller that the limits overlap with similar strength.
In this crucial region for both subcombinations, the full combination then
improves the sensitivity by up to 50\% at 4~\TeV.
 
\begin{figure}[b]
\centering
\subfloat[]{\includegraphics[width=0.49\textwidth]{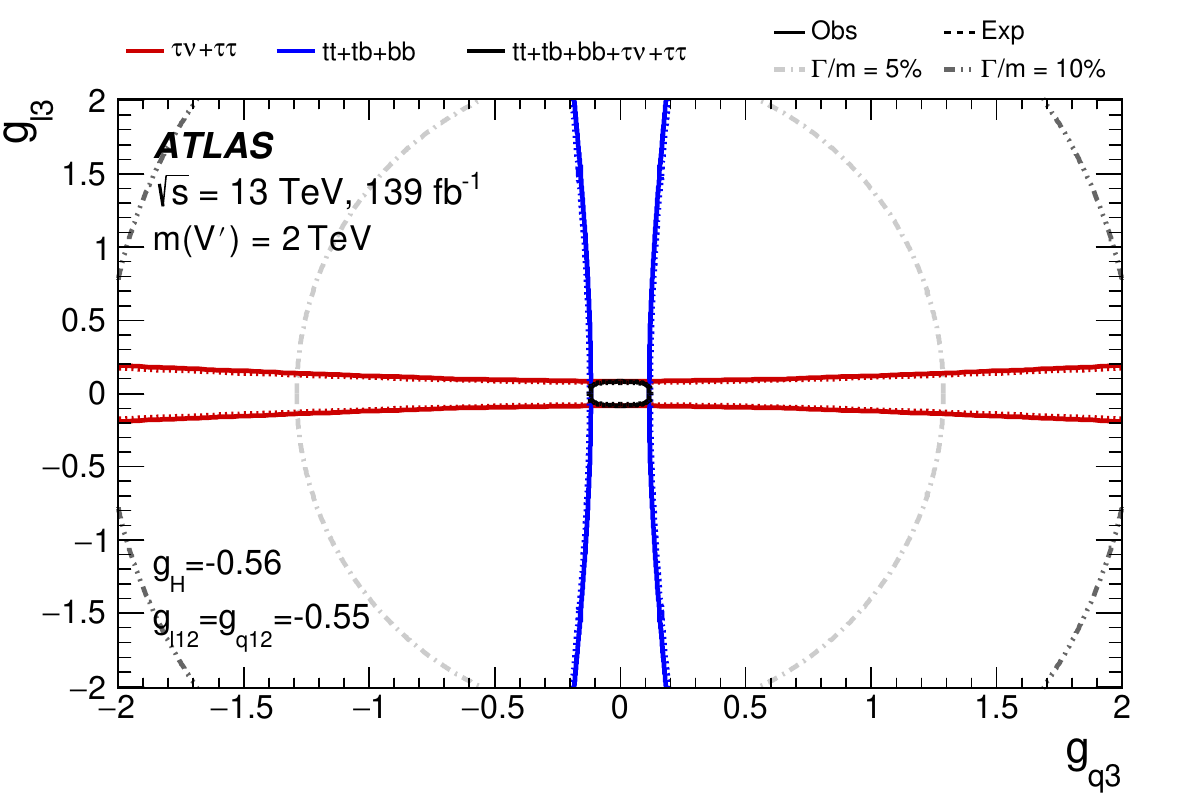}}
\subfloat[]{\includegraphics[width=0.49\textwidth]{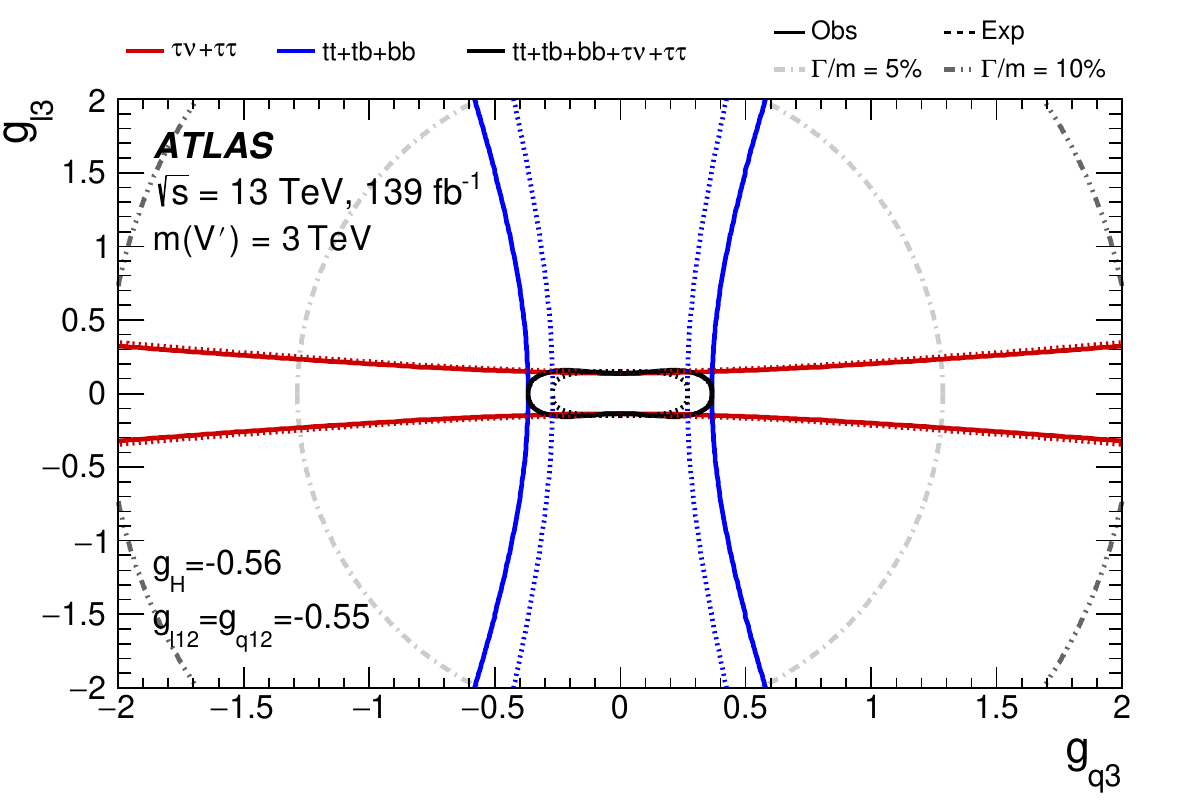}}\\
\subfloat[]{\includegraphics[width=0.49\textwidth]{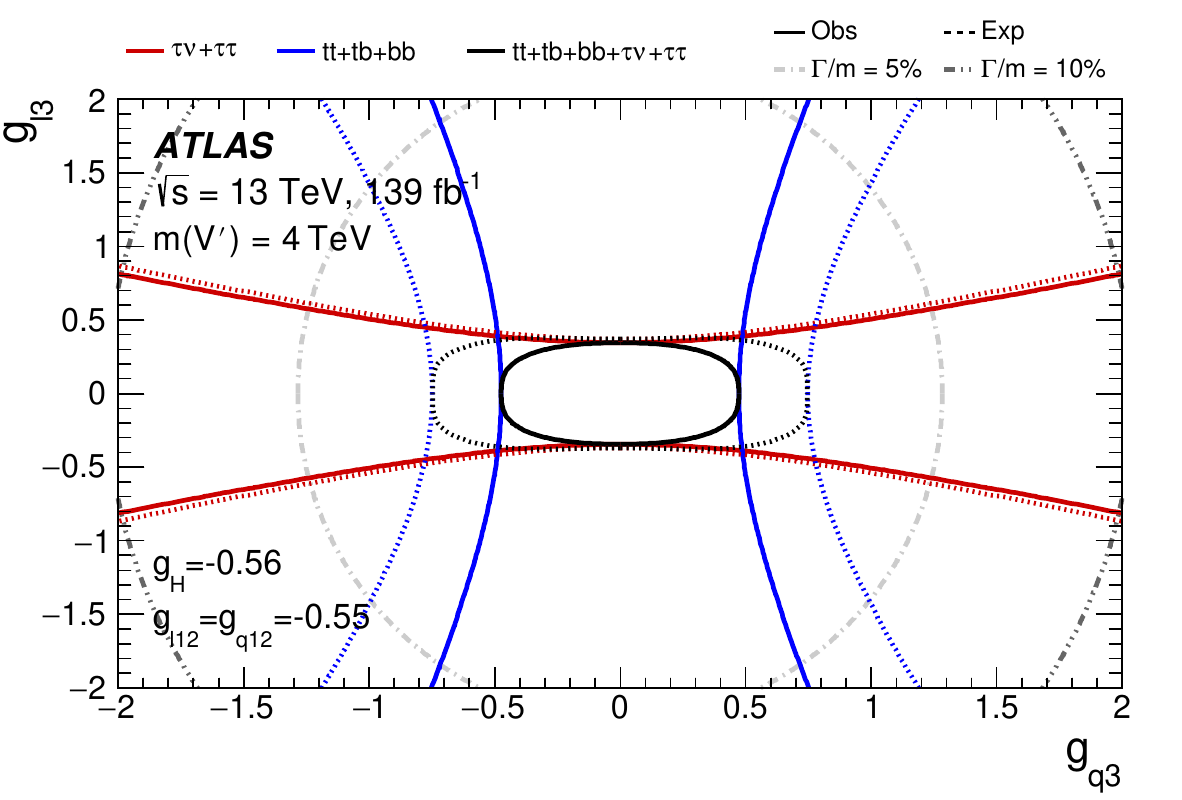}}
\caption{95\% CL observed and expected upper limit contours for the full combination in
the 2D coupling plane for $\{g_{q3}, g_{\ell_3}\}$ at a signal pole mass of (a) 2~\TeV,
(b) 3~\TeV, and (c) 4~\TeV\ for the $q\bar{q}$ production mode. The dashed
grey lines show the region where the resonance natural width is either 5\% or 10\% of the pole mass.}
\label{fig:DY_2D_gQ3gL3_Comb}
\end{figure}
 
The conclusions are very similar for the $\{g_{q_3}, g_{\ell_3}\}$ plane, shown in Figure~\ref{fig:DY_2D_gQ3gL3_Comb}.
The third-generation quark and lepton channels dominate their own regions of phase space,
but the full combination brings large improvements, especially at high mass.
While a small excess in data is observed for the third-generation quark channels at 2~\TeV,
and a deficit at 3~\TeV, no such equivalent excess or deficit is observed in the
third-generation lepton channels.

% End of text imported from the .//sections/9-results.tex input file

\FloatBarrier

% The next lines are included from the .//sections/10-conclusion.tex input file
\section{Conclusion}
 
A combination of results from searches for heavy-resonance production in various bosonic, quarkonic, and
leptonic final states is presented. The data were collected with the ATLAS detector at the LHC in \pp collisions at
$\rts = 13$~\TeV\ and correspond to an integrated luminosity of 139~\ifb.
While previous combination efforts included only the decays of heavy
resonances into bosonic, and some leptonic, final states, the combination presented here also includes decays into quarkonic
final states, as well as third-generation fermion final states for the first time. Compared to the individual
analyses, the combined results strengthen the constraints on physics beyond the Standard Model and allow the constraints to be
expressed in terms of the couplings to quarks, leptons, or bosons. The sensitivities of the different production
mechanisms are compared, such as quark--antiquark annihilation versus vector-boson fusion.
 
The combined results are interpreted in the context of models with a heavy vector-boson triplet. As no signficant excess of events is observed in the data, exclusion limits are set in these models. The observed (expected) 95\% CL lower limit on the mass of $V^{\prime}$ resonances in the weakly coupled HVT model A is 5.8 (5.6)~\TeV, and the corresponding limit in the strongly coupled HVT model B is 4.4 (4.4)~\TeV. For the VBF production mode (HVT model C benchmark point) the corresponding limit is 0.4 (0.5)~\TeV, while for the cases where $g_H = 3$ or $g_H = 6$, the observed (expected) limits are 1.0 (1.1)~\TeV\ and 1.5 (1.8)~\TeV, respectively. The combined results are used to place stringent constraints on couplings of heavy vector bosons to quarks, leptons, and bosons. These constraints improve upon previous limits in the coupling space by up to 60\% depending on the resonance mass and the specific
coupling parameters.

% End of text imported from the .//sections/10-conclusion.tex input file

\section*{Acknowledgements}

% The next lines are included from the .//acknowledgements/Acknowledgements.tex input file

We thank CERN for the very successful operation of the LHC and its injectors, as well as the support staff at
CERN and at our institutions worldwide without whom ATLAS could not be operated efficiently.
 
The crucial computing support from all WLCG partners is acknowledged gratefully, in particular from CERN, the ATLAS Tier-1 facilities at TRIUMF/SFU (Canada), NDGF (Denmark, Norway, Sweden), CC-IN2P3 (France), KIT/GridKA (Germany), INFN-CNAF (Italy), NL-T1 (Netherlands), PIC (Spain), RAL (UK) and BNL (USA), the Tier-2 facilities worldwide and large non-WLCG resource providers. Major contributors of computing resources are listed in Ref.~\cite{ATL-SOFT-PUB-2023-001}.
 
We gratefully acknowledge the support of ANPCyT, Argentina; YerPhI, Armenia; ARC, Australia; BMWFW and FWF, Austria; ANAS, Azerbaijan; CNPq and FAPESP, Brazil; NSERC, NRC and CFI, Canada; CERN; ANID, Chile; CAS, MOST and NSFC, China; Minciencias, Colombia; MEYS CR, Czech Republic; DNRF and DNSRC, Denmark; IN2P3-CNRS and CEA-DRF/IRFU, France; SRNSFG, Georgia; BMBF, HGF and MPG, Germany; GSRI, Greece; RGC and Hong Kong SAR, China; ISF and Benoziyo Center, Israel; INFN, Italy; MEXT and JSPS, Japan; CNRST, Morocco; NWO, Netherlands; RCN, Norway; MEiN, Poland; FCT, Portugal; MNE/IFA, Romania; MESTD, Serbia; MSSR, Slovakia; ARRS and MIZ\v{S}, Slovenia; DSI/NRF, South Africa; MICINN, Spain; SRC and Wallenberg Foundation, Sweden; SERI, SNSF and Cantons of Bern and Geneva, Switzerland; MOST, Taipei; TENMAK, T\"urkiye; STFC, United Kingdom; DOE and NSF, United States of America.
 
Individual groups and members have received support from BCKDF, CANARIE, CRC and DRAC, Canada; PRIMUS 21/SCI/017 and UNCE SCI/013, Czech Republic; COST, ERC, ERDF, Horizon 2020, ICSC-NextGenerationEU and Marie Sk{\l}odowska-Curie Actions, European Union; Investissements d'Avenir Labex, Investissements d'Avenir Idex and ANR, France; DFG and AvH Foundation, Germany; Herakleitos, Thales and Aristeia programmes co-financed by EU-ESF and the Greek NSRF, Greece; BSF-NSF and MINERVA, Israel; Norwegian Financial Mechanism 2014-2021, Norway; NCN and NAWA, Poland; La Caixa Banking Foundation, CERCA Programme Generalitat de Catalunya and PROMETEO and GenT Programmes Generalitat Valenciana, Spain; G\"{o}ran Gustafssons Stiftelse, Sweden; The Royal Society and Leverhulme Trust, United Kingdom.
 
In addition, individual members wish to acknowledge support from CERN: European Organization for Nuclear Research (CERN PJAS); Chile: Agencia Nacional de Investigación y Desarrollo (FONDECYT 1190886, FONDECYT 1210400, FONDECYT 1230987); China: National Natural Science Foundation of China (NSFC - 12175119, NSFC 12275265); European Union: European Research Council (ERC - 948254, ERC 101089007), Horizon 2020 Framework Programme (MUCCA - CHIST-ERA-19-XAI-00), Italian Center for High Performance Computing, Big Data and Quantum Computing (ICSC, NextGenerationEU); France: Agence Nationale de la Recherche (ANR-20-CE31-0013, ANR-21-CE31-0022), Investissements d'Avenir Labex (ANR-11-LABX-0012); Germany: Baden-Württemberg Stiftung (BW Stiftung-Postdoc Eliteprogramme), Deutsche Forschungsgemeinschaft (DFG - 469666862, DFG - CR 312/5-2); Italy: Istituto Nazionale di Fisica Nucleare (ICSC, NextGenerationEU); Japan: Japan Society for the Promotion of Science (JSPS KAKENHI 22H01227, JSPS KAKENHI 22KK0227, JSPS KAKENHI JP21H05085, JSPS KAKENHI JP22H04944); Netherlands: Netherlands Organisation for Scientific Research (NWO Veni 2020 - VI.Veni.202.179); Norway: Research Council of Norway (RCN-314472); Poland: Polish National Agency for Academic Exchange (PPN/PPO/2020/1/00002/U/00001), Polish National Science Centre (NCN 2021/42/E/ST2/00350, NCN OPUS nr 2022/47/B/ST2/03059, NCN UMO-2019/34/E/ST2/00393, UMO-2020/37/B/ST2/01043, UMO-2021/40/C/ST2/00187, UMO-2022/47/O/ST2/00148); Slovenia: Slovenian Research Agency (ARIS grant J1-3010); Spain: BBVA Foundation (LEO22-1-603), Generalitat Valenciana (Artemisa, FEDER, IDIFEDER/2018/048), Ministry of Science and Innovation (RYC2019-028510-I, RYC2020-030254-I), PROMETEO and GenT Programmes Generalitat Valenciana (CIDEGENT/2019/023, CIDEGENT/2019/027); Sweden: Swedish Research Council (VR 2022-03845), Knut and Alice Wallenberg Foundation (KAW 2022.0358); Switzerland: Swiss National Science Foundation (SNSF - PCEFP2\_194658); United Kingdom: Leverhulme Trust (Leverhulme Trust RPG-2020-004); United States of America: Neubauer Family Foundation.

% End of text imported from the .//acknowledgements/Acknowledgements.tex input file

\printbibliography
 
\clearpage
\input{atlas_authlist}

\end{document}

%% file: atlas_authlist.tex
% ATLAS Collaboration author list
% Reference date of EXOT-2022-38 is 2023-12-01
% Author list last updated on date 25-MAR-24
% Data extracted on 25-Mar-2024 for paper reference EXOT-2022-38
% at 9:05pm
 
\begin{flushleft}
\hypersetup{urlcolor=black}
{\Large The ATLAS Collaboration}

\bigskip

\AtlasOrcid[0000-0002-6665-4934]{G.~Aad}$^\textrm{\scriptsize 103}$,
\AtlasOrcid[0000-0001-7616-1554]{E.~Aakvaag}$^\textrm{\scriptsize 16}$,
\AtlasOrcid[0000-0002-5888-2734]{B.~Abbott}$^\textrm{\scriptsize 121}$,
\AtlasOrcid[0000-0002-1002-1652]{K.~Abeling}$^\textrm{\scriptsize 55}$,
\AtlasOrcid[0000-0001-5763-2760]{N.J.~Abicht}$^\textrm{\scriptsize 49}$,
\AtlasOrcid[0000-0002-8496-9294]{S.H.~Abidi}$^\textrm{\scriptsize 29}$,
\AtlasOrcid[0009-0003-6578-220X]{M.~Aboelela}$^\textrm{\scriptsize 44}$,
\AtlasOrcid[0000-0002-9987-2292]{A.~Aboulhorma}$^\textrm{\scriptsize 35e}$,
\AtlasOrcid[0000-0001-5329-6640]{H.~Abramowicz}$^\textrm{\scriptsize 152}$,
\AtlasOrcid[0000-0002-1599-2896]{H.~Abreu}$^\textrm{\scriptsize 151}$
% ...
\bigskip
\\

% ...
$^{16}$Department for Physics and Technology, University of Bergen, Bergen; Norway.\\
% ...
$^{29}$Physics Department, Brookhaven National Laboratory, Upton NY; United States of America.\\
% ...
$^{35}$$^{(a)}$Facult\'e des Sciences Ain Chock, R\'eseau Universitaire de Physique des Hautes Energies - Universit\'e Hassan II, Casablanca;$^{(b)}$Facult\'{e} des Sciences, Universit\'{e} Ibn-Tofail, K\'{e}nitra;$^{(c)}$Facult\'e des Sciences Semlalia, Universit\'e Cadi Ayyad, LPHEA-Marrakech;$^{(d)}$LPMR, Facult\'e des Sciences, Universit\'e Mohamed Premier, Oujda;$^{(e)}$Facult\'e des sciences, Universit\'e Mohammed V, Rabat;$^{(f)}$Institute of Applied Physics, Mohammed VI Polytechnic University, Ben Guerir; Morocco.\\
% ...
$^{44}$Physics Department, Southern Methodist University, Dallas TX; United States of America.\\
% ...
$^{49}$Fakult\"{a}t Physik , Technische Universit{\"a}t Dortmund, Dortmund; Germany.\\
% ...
$^{55}$II. Physikalisches Institut, Georg-August-Universit\"{a}t G\"ottingen, G\"ottingen; Germany.\\
% ...
$^{103}$CPPM, Aix-Marseille Universit\'e, CNRS/IN2P3, Marseille; France.\\
% ...
$^{121}$Homer L. Dodge Department of Physics and Astronomy, University of Oklahoma, Norman OK; United States of America.\\
% ...
$^{151}$Department of Physics, Technion, Israel Institute of Technology, Haifa; Israel.\\
$^{152}$Raymond and Beverly Sackler School of Physics and Astronomy, Tel Aviv University, Tel Aviv; Israel.\\
% ...
$^{a}$ Also Affiliated with an institute covered by a cooperation agreement with CERN.\\
$^{b}$ Also at An-Najah National University, Nablus; Palestine.\\
$^{c}$ Also at Borough of Manhattan Community College, City University of New York, New York NY; United States of America.\\
$^{d}$ Also at Center for High Energy Physics, Peking University; China.\\
$^{e}$ Also at Center for Interdisciplinary Research and Innovation (CIRI-AUTH), Thessaloniki; Greece.\\
$^{f}$ Also at Centro Studi e Ricerche Enrico Fermi; Italy.\\
% ...

\end{flushleft}

% Created with Glance <Atlas.Glance@cern.ch>